\documentclass{JHEP3}
\usepackage{graphics}

\def\half{{1\over 2}}
\def\tilde{\widetilde}

\def\Id{{\mathbb{I}}}

\def\ah{{\widehat\alpha}}

\def\bh{{\widehat\beta}}
\def\gh{{\widehat\gamma}}

\def\Str{{\textrm{Str}}}
\def\a{\alpha}
\def\b{\beta}
\def\g{\gamma}
\def\d{\delta}
\def\e{\epsilon}

\def\l{\lambda}

\def\t{{\theta}}

\def\ah{{\hat \alpha}}

\def\ad{{\dot \a}}
\def\bd{{\dot \b}}

\def\nn{{\nonumber}}
\def\ZZ{\ensuremath{\mathbb{Z}}}

\newcommand{\QQ}{{\mbox{\small \bf Q}}}

\newcommand{\HH}{{\mbox{\small \bf H}}}

\newcommand{\EE}{{\mbox{\small \bf E}}}

\newcommand{\be}{\begin{equation}}
\newcommand{\ee}{\end{equation}}
\newcommand{\bea}{\begin{eqnarray}}
\newcommand{\eea}{\end{eqnarray}}

\title{Integrability of Type II Superstrings on Ramond-Ramond
Backgrounds in Various Dimensions}

\author{Ido Adam, Amit Dekel, Luca Mazzucato
  and Yaron Oz\\
Raymond and Beverly Sackler Faculty of Exact Sciences \\
School of Physics and Astronomy \\
Tel-Aviv University, Ramat-Aviv 69978, Israel\\
E-mails: \email{adamido@post.tau.ac.il},
  \email{amitde@post.tau.ac.il}, \email{mazzul@post.tau.ac.il},
  \email{yaronoz@post.tau.ac.il}}

\abstract{We consider type II superstrings on AdS backgrounds with
Ramond-Ramond flux in various dimensions. We realize the backgrounds
as supercosets and analyze explicitly two classes of models:
non-critical superstrings on $AdS_{2d}$ and critical superstrings on
$AdS_p\times S^p\times CY$. We work both in the Green--Schwarz and in
the pure spinor formalisms. We construct a one-parameter family of
flat currents (a Lax connection), leading to an infinite number of
conserved non-local charges, which imply the classical integrability
of both sigma-models. In the pure spinor formulation, we use the BRST
symmetry to prove the quantum integrability of the sigma-model. We
discuss how classical $\kappa$-symmetry implies one-loop conformal
invariance. We consider the addition of space-filling D-branes to the
pure spinor formalism.}


\preprint{TAUP-2849/07}

\begin{document}

\section{Introduction and summary}

Superstring theory on $AdS$ backgrounds with Ramond-Ramond flux has
not been quantized yet. The Green-Schwarz sigma-model on such
backgrounds is an interacting two-dimensional conformal field
theory. In the case of the type IIB superstring on $AdS_5\times S^5$
\cite{Metsaev:1998it}, the authors of \cite{Bena:2003wd} showed that
the sigma-model is invariant under a Yangian symmetry algebra and as a
result is classically integrable. Their result relies on the
realization of the background as a supercoset $G/H$, where $G$ is a
supergroup with a $\mathbb{Z}_4$ automorphism group and $H$ is the
$\mathbb{Z}_4$ fixed locus bosonic subgroup of $G$. Once uncovering
this hidden symmetry, one can ask whether the Yangian algebra, derived
for the $AdS_5\times S^5$ background, is a general feature of
superstrings on $AdS$ backgrounds with RR flux.

We will address this question by looking at superstring theories on
such backgrounds, both in the Green--Schwarz and the pure spinor
formalisms. We will first construct sigma-model actions and find
simple actions for the Green-Schwarz and the pure spinor superstrings,
which hold in all dimensions. We will then show classical
integrability of both sigma-models as well as quantum integrability of
the pure spinor one.\footnote{Recently, these kinds of supercoset
sigma-models have received attention regarding their integrability
properties, see for example
\cite{Polyakov:2004br,Polyakov:2005ss,Chen:2005uj,Kagan:2005wt,Young:2005jv,Kluson:2006ij,Babichenko:2006uc}.}

In general for the GS superstring, it is difficult to analyze the
quantum sigma-model. This is because the quantization of the GS
superstring is known only in the light-cone gauge and hence
non-covariantly. Since the equations of motion of the GS superstring
do not provide a propagator for the $\theta$'s, the calculations in
worldsheet perturbation theory are problematic. On the other hand, the
pure spinor sigma-model can be quantized in a straightforward manner,
since it contains additional terms that break explicitly the GS
$\kappa$-symmetry and introduce propagators for all the
variables. Hence, we will be able to show that our models are gauge
invariant and BRST invariant at all orders in the worldsheet
perturbation theory using the methods of \cite{Berkovits:2004xu}.

We will consider explicitly two classes of models: Type II
non-critical superstrings on $AdS_{2d}$, for
 $d=1,2,3$, and Type II critical superstrings on $AdS_p\times
S^p\times CY_{5-p}$, for $p=2,3$.

The first class of models are strongly coupled two-dimensional
CFTs. The sigma-model coupling, given by the curvature of $AdS$, is
fixed to a finite value of order one in string units, and the theory
cannot be analyzed perturbatively. The worldsheet variables for the
non-critical superstrings and in particular their pure spinor spaces
have been derived in \cite{Adam:2006bt} by mapping the RNS formulation
of the linear dilaton background to the covariant one.\footnote{See
also \cite{Grassi:2005kc} for the hybrid formulation of the linear
dilaton background and
\cite{Wyllard:2005fh,Grassi:2005sb,Kunitomo:2006gr} about
lower-dimensional pure spinor superstrings.}

In the second class of models, the sigma-model will describe the
non-compact part $AdS_p\times S^p$ of critical superstrings on
ten-dimensional backgrounds. Unlike the previous non-critical string
case, the curvature of $AdS$ is a modulus. Thus, one can take the
limit in which the curvature is small and the sigma-model is weakly
coupled and can be studied perturbatively.

All our models are realized as nonlinear sigma-models on supercosets
$G/H$, where the supergroup $G$ has a $\ZZ_4$ automorphism, whose
invariant locus is $H$. A crucial property of sigma-models on such
supercosets is their classical integrability. In order to exhibit the
integrability of the sigma-models, we have to construct an infinite
number of conserved charges
\cite{Young:2005jv,Chen:2005uj}. Furthermore, for the charges to be
physical they have to be $\kappa$-invariant and BRST invariant in the
Green--Schwarz and in the pure spinor formalisms, respectively. The
first step in the construction of the charges is to find a
one-parameter family of currents $a(\mu)$ satisfying the flatness
condition
\begin{equation}
da(\mu) + a(\mu) \wedge a(\mu) = 0 \ .\label{laxin}
\end{equation}
One then constructs the Wilson line \be U_{(\mu)}(x,t;y,t) =
\mathrm{P} \exp \left( -\int_{(y,t)}^{(x,t)}a(\mu) \right) \ ,
\label{wilson} \ee and obtains the infinite set of non-local
charges $Q_n$ by expanding \be U_{(\mu)}(\infty,t;-\infty,t) = 1+
\sum_{n=1}^{\infty} \mu^n Q_n \ . \label{nlocal} \ee The conservation
of $Q_n$ is implied by the flatness of $a(\mu)$ (provided $a(\mu)$
vanishes at $\pm \infty$).\footnote{Establishing the existence of the
Lax connection (\ref{laxin}) is the first step towards the solution of
the nonlinear sigma-model. In particular, this technique has been
fully exploited in \cite{Beisert:2005bm} to find the classical
spectrum of the GS type IIB superstring on $AdS_5\times S^5$.}  This
is valid for a sigma model on a plane. In the closed string case, we
need to impose periodic boundary conditions and hence consider a
slightly different invariant --- the trace of the Wilson loop.

The first two charges $Q_1$ and $Q_2$ generate the Yangian
algebra, which is a symmetry algebra underlying the type II
superstrings propagating on the AdS backgrounds with Ramond-Ramond
fluxes in various dimensions. Moreover, in the pure spinor
formalism one can see that this symmetry holds also at the quantum
sigma-model level. This has been shown by Berkovits in the
$AdS_5\times S^5$ background in \cite{Berkovits:2004xu}. We
will show that quantum integrability of the pure spinor action
holds also in the lower-dimensional cases. In the case of type IIB
superstrings propagating on $AdS_5\times S^5$, a similar Yangian
algebra has been identified in the free field theory limit of
$\mathcal{N}=4$ SYM at large $N_c$ \cite{Dolan:2003uh}. We expect that a
similar structure underlies the field theory duals in various
dimensions.

Note, that the Yangian algebra suggests the existence of an affine
Kac-Moody algebra \cite{Dolan:1981fq}. This is to be contrasted with
NS-NS backgrounds, where the affine algebra comes in two copies, one
left- and one right-moving, while in the case of RR backgrounds there
would be only a single copy of such an algebra.  The question arises
whether this symmetry is sufficient for solving for the spectrum of the
superstring.

The paper is organized as follows. Throughout most of the paper, we
analyze in general the structure of superstrings sigma-models on
supercosets with $\ZZ_4$ automorphisms. As we will see, most of their
properties are algebraic and do not rely on the particular choice of
the supercoset. In section 2 we introduce the classical
$\kappa$-invariant Green-Schwarz sigma-model and find a one-parameter
family of flat currents (Lax connection). This leads to an infinite
number of conserved non-local charges and shows classical
integrability of the sigma-model. In section 3 we introduce the pure
spinor action and compute the one-parameter family of flat currents,
which is different from the GS one. We describe also the various pure
spinor spaces that we use in the various dimensions. At the end of the
section, we discuss the addition of open string boundary conditions to
the pure spinor sigma-model. In section 4, we study the pure spinor
sigma-model at the quantum level and show that it is gauge invariant
and BRST invariant at all orders in perturbation theory and argue that
for $AdS_2$ these properties hold non-perturbatively as well. By using
BRST symmetry, we then prove quantum integrability. In section 5, we
study one-loop conformal invariance of the GS sigma-model and then
describe the various specific backgrounds and their supercoset
realizations in section 6. In Appendices A and B we collected some
technical details of the GS and pure spinor computations, while in
Appendix C we describe the various supergroups and their notations. In
Appendix D we review the supergravity solution of non-critical
$AdS_5\times S^1$ of \cite{Klebanov:2004ya} and find a curious result about
the higher curvature corrections to this solution.

\section{Integrability of  Green-Schwarz superstrings on RR backgrounds}

In this section we will consider the integrability properties of
Green-Schwarz superstrings on the background of a supercoset $G/H$
with only RR-flux, where $G$ is a supergroup with a $\mathbb{Z}_4$
automorphism whose invariant locus is the subgroup $H$. We will
construct the Green-Schwarz action and derive the family of flat
connections leading to an infinite number of conserved non-local
charges\footnote{Our construction will be covariant. In the case of
$AdS_5\times S^5$, it has been shown in \cite{Alday:2005gi} that the
Green-Schwarz sigma-model is still integrable after gauge fixing of
$\kappa$-symmetry and reparametrization invariance.}. The
$\kappa$-invariance of the currents in the GS formalism will follow
from the BRST invariance of the non-local currents in the pure spinor
formalism that we will prove in the next section as explained in
\cite{Berkovits:2004jw}. The notations about supergroups are
summarized in the appendix. The discussion in this section will be at
a formal level, while we will specialize to the particular backgrounds
in sections \ref{sec:one-loop-conformal-inv} and \ref{examples}.

\subsection{The Green-Schwarz sigma-model}
\label{gssm}

We will be interested in sigma-models whose target space is the
coset $G/H$, where $G$ is a supergroup with a $\mathbb{Z}_4$
automorphism and the subgroup $H$ is the invariant locus of this
automorphism. The super Lie algebra $\mathcal{G}$ of $G$ can be
decomposed into the $\mathbb{Z}_4$ automorphism invariant spaces
$\mathcal{G} = \mathcal{H}_0 \oplus \mathcal{H}_1 \oplus
\mathcal{H}_2 \oplus \mathcal{H}_3$, where the subscript keeps
track of the $\mathbb{Z}_4$ charge and in particular
$\mathcal{H}_0$ is the algebra of the subgroup $H$. This
decomposition satisfies the algebra $(i=1, \dots, 3)$
\begin{equation}
  [\mathcal{H}_0, \mathcal{H}_0] \subset \mathcal{H}_0 \ , \quad
  [\mathcal{H}_0, \mathcal{H}_i] \subset \mathcal{H}_i \ , \quad
  [\mathcal{H}_i, \mathcal{H}_j] \subset \mathcal{H}_{i + j \
  \mathrm{mod}\ 4} .
\end{equation}
and the only non-vanishing supertraces%
\footnote{The supertrace of a supermatrix $M = \left(
\begin{array}{cc}
  A & B \\
  C & D
\end{array}
\right)$ is defined as $\langle M \rangle = \mathrm{Str} M =
\mathrm{tr} A - (-1)^{\mathrm{deg} M} \mathrm{tr} D$, where
$\mathrm{deg} M$ is $0$ for Grassmann even matrices and $1$ for
Grassmann odd ones. } are
\begin{equation}
  \langle \mathcal{H}_i \mathcal{H}_j \rangle \neq 0 \ ,\  i + j = 0 \
  \mathrm{mod}\ 4 \quad (i, j = 0,
  \dots, 3) \ .
\end{equation}
We will denote the bosonic generators in $\mathcal{G}$ by $T_{[a
b]} \in \mathcal{H}_0$, $T_a \in \mathcal{H}_2$, and the fermionic
ones by $T_\alpha \in \mathcal{H}_1$, $T_{\hat \alpha} \in
\mathcal{H}_3$.

The worldsheet fields are the maps $g: \Sigma \to G$ and dividing
by the subgroup $H$ is done by gauging the subgroup $H$ acting
from the right by $g \simeq g h$, $h \in H$. The sigma-model is
further constrained by the requirement that it be invariant under
the global symmetry $g \to \hat g g$, $\hat g \in G$. The
left-invariant current is  defined as
\begin{equation}
  J = g^{-1} d g \ ,
\end{equation}
which satisfies the Maurer-Cartan equation
\begin{equation}
  d J + J \wedge J = 0 \ .
\end{equation}
This current can be decomposed according to the $\mathbb{Z}_4$
grading of the algebra $J = J_0 + J_1 + J_2 + J_3$ and the
Maurer-Cartan equation splits into
\begin{eqnarray}
  d J_0 + J_0 \wedge J_0 + J_1 \wedge J_3 + J_2 \wedge J_2 + J_3
  \wedge J_1 & = & 0 \ ,\label{eq:Maurer-Cartan1} \\
  d J_1 + J_0 \wedge J_1 + J_1 \wedge J_0 + J_2 \wedge J_3 + J_3
  \wedge J_2 & = & 0 \ , \\
  d J_2 + J_0 \wedge J_2 + J_1 \wedge J_1 + J_2 \wedge J_0 + J_3
  \wedge J_3 & = & 0 \ , \\
  d J_3 + J_0 \wedge J_3 + J_1 \wedge J_2 + J_2 \wedge J_1 + J_3
  \wedge J_0 & = & 0 \ . \label{eq:Maurer-Cartan4}
\end{eqnarray}
These currents are manifestly invariant under the global symmetry,
which acts by left multiplication. Under the gauge transformation,
which acts by right multiplication, they transform as
\begin{equation}
  \delta J = d \Lambda + [J, \Lambda] \ , \quad \Lambda \in
  \mathcal{H}_0 \ .
\end{equation}

Using the above properties of the algebra $\mathcal{G}$ and the
requirement of gauge invariance leads to the GS action (in the
following we will use $J_i$ both to denote the $1$-form currents
in the target space as well as their pullback to the worldsheet)
\begin{equation} \label{eq:generic-GS-coset-action}
  S_\mathrm{GS} = \frac{1}{4} \int \langle J_2 \wedge *J_2 + J_1
  \wedge J_3 \rangle = \frac{1}{4} \int d^2 \sigma \langle \sqrt{h}
  h^{m n} J_{2 m} J_{2 n} + \epsilon^{m n} J_{1 m} J_{3 n} \rangle \ ,
  \label{eq:coset-GS-action}
\end{equation}
where $m, n = 1, 2$ are worldsheet indices.  A $J_0 \wedge *J_0$
term does not appear because of gauge invariance, while the term
$J_1 \wedge *J_3$ breaks $\kappa$-symmetry and therefore cannot be
included in the GS action. The first and second terms in the
action are the kinetic and Wess-Zumino terms, respectively. The
coefficient of the Wess-Zumino term is determined using
$\kappa$-symmetry as shown in the next paragraph. For a particular
choice of the supergroup, this GS action reproduces the GS action
on $AdS_2$ background constructed in \cite{Verlinde:2004gt} and
the GS action on $AdS_5 \times S^5$ \cite{Metsaev:1998it}.

Let us verify now that the action is indeed invariant under
$\kappa$-symmetry. It is convenient to parameterize the
$\kappa$-transformation by \cite{Metsaev:1998it}
\begin{equation}
  \delta_\kappa x_i \equiv \delta_\kappa X^M J_{i M} \;\;  \ ,
\end{equation}
where the index $M$ runs over the target superspace indices and
$X^M$ are the superspace coordinates, while $i = 1,
  \dots, 3$ denotes the $\mathbb{Z}_4$ grading. Since $J_i = d X^M
J_{i M}$ we obtain the following transformations of the currents
\begin{eqnarray}
  \delta_\kappa J_2 & = & d \delta_\kappa x_2 + [J_0, \delta_\kappa x_2] +
  [J_2, \delta_\kappa x_0] + [J_1, \delta_\kappa x_1] + [J_3,
  \delta_\kappa x_3] \ , \\
  \delta_\kappa J_1 & = & d \delta_\kappa x_1 + [J_0, \delta_\kappa
  x_1] + [J_1, \delta_\kappa x_0] + [J_2, \delta_\kappa x_3] + [J_3,
  \delta_\kappa x_2] \ , \\
  \delta_\kappa J_0 & = & d \delta_\kappa x_0 + [J_0, \delta_\kappa
  x_0] + [J_1, \delta_\kappa x_3] + [J_2, \delta_\kappa x_2] + [J_3,
  \delta_\kappa x_1] \ , \\
  \delta_\kappa J_3 & = & d \delta_\kappa x_3 + [J_0, \delta_\kappa
  x_3] + [J_1, \delta_\kappa x_2] + [J_2, \delta_\kappa x_1] + [J_3,
  \delta_\kappa x_0] \ .
\end{eqnarray}
Using these transformations and taking into account the
Maurer-Cartan equations, the $\kappa$-transformation of the
actions is
\begin{eqnarray}
  \delta_\kappa S_\mathrm{GS} & = & \frac{1}{4} \int d^2 \sigma
  \langle \epsilon^{m n} \partial_m (J_{3 n} \delta_\kappa x_1 -
  J_{1 n} \delta_\kappa x_3) + \delta_\kappa (\sqrt{h}
  h^{m n}) J_{2 m} J_{2 n} + \nonumber \\
  && {} +2 \sqrt{h} h^{m n} (J_{2 m}
  \partial_n \delta_\kappa x_2 + [J_{2 m}, J_{0 n}] \delta_\kappa x_2)
  + \epsilon^{m n} ([J_{1 m}, J_{1 n}] - [J_{3 m}, J_{3 n}])
  \delta_\kappa x_2 - \nonumber \\
  && {} - 2 (\sqrt{h} h^{m n} + \epsilon^{m n}) [J_{1 n},
  J_{2 m}] \delta_\kappa x_1 + 2 (\sqrt{h} h^{m n} - \epsilon^{m n})
  [J_{2 m}, J_{3 n}] \delta_\kappa x_3 \rangle \ .
\end{eqnarray}
The $\kappa$-transformation is parameterized by
\begin{equation}
  \delta_\kappa x_2 = 0 \ , \quad \delta_\kappa x_1 = [J_{2 m},
  \kappa_3^m] \ , \quad \delta_\kappa x_3 = [J_{2 m}, \kappa_1^m] \ ,
  \label{kappasym}
\end{equation}
where $\kappa_3^m \in \mathcal{H}_3$ and $\kappa_1^m \in
\mathcal{H}_1$. By substituting this and expressing the result in
terms of the structure constants and the Cartan metric $\eta$ one
finally has
\begin{eqnarray}
  \delta_\kappa S_\mathrm{GS} & = & \frac{1}{4} \int d^2 \sigma \Big[
  \epsilon^{m n} \langle \partial_m (J_{3 n} \delta_\kappa x_1 - J_{1
  n} \delta_\kappa x_3) \rangle + \delta_\kappa (\sqrt{h} h^{m n})
  \eta_{a b} J_{2 m}^a J_{2 n}^b + \nonumber \\
  && {} + 4 \sqrt{h} (P_+^{m n} \eta_{\hat \beta \beta} f_{\alpha a}^{\hat
  \beta} f_{b \hat \alpha}^\beta J_{1 n}^\alpha \kappa_3^{p \hat
  \alpha} -  P_-^{m n} \eta_{\beta \hat \beta} f_{a \hat \alpha}^\beta
  f_{b \alpha}^{\hat \beta} J_{3 n}^{\hat \alpha} \kappa_1^{p \alpha})
  J_{2 m}^a J_{2 p}^b \Big] \ ,
\end{eqnarray}
where we have defined the projectors $P_\pm^{m n} =
\frac{1}{2}(h^{m n} \pm \frac{1}{\sqrt{h}} \epsilon^{m n})$. Since
$\delta_\kappa (\sqrt{h} h^{m n})$ should be symmetric and
traceless and not Lie-algebra valued, we have to require that
\begin{equation} \label{eq:kappa-symmetry-condition}
   \eta_{\beta \hat \beta} \left( f_{a \hat \alpha}^\beta
  f_{b \alpha}^{\hat \beta} + f_{b \hat \alpha}^\beta f_{a
  \alpha}^{\hat \beta} \right) = c_{\alpha \hat \alpha} \eta_{a b}
\end{equation}
for some matrix $c_{\alpha \hat \alpha}$. Then one obtains
\begin{equation}
  \delta_\kappa (\sqrt{h} h^{m n}) = 4 \sqrt{h} c_{\alpha \hat \alpha}
  (P_-^{m p} J_{3 p}^{\hat \alpha} \kappa_1^{n \alpha} - P_+^{m p}
  J_{1 p}^\alpha k_3^{n \hat \alpha}) \ ,
\end{equation}
which is automatically symmetric in $a$ and $b$ if we require that
\begin{equation}
  \kappa_1^m = P_-^{m n} \kappa_{1 n} \ , \quad \kappa_3^m = P_+^{m n}
  \kappa_{3 n}
\end{equation}
since $P_\pm^{m p} P_\pm^{n q} = P_\pm^{n p} P_\pm^{m q}$. It is
also traceless because $P_-^{n m} \kappa_{1 n} = P_+^{n m}
\kappa_{3 n} = 0$.

The relation (\ref{eq:kappa-symmetry-condition}), required for
$\kappa$-symmetry, is a condition on the structure constants of the
supergroup. This condition is equivalent to the torsion constraints of
type II supergravity in various dimensions.\footnote{By ``type II"
supergravity in dimension $D$ we mean a theory with as many gravitini
as the ones we would get by compactifying ten-dimensional type II
supergravity on a Calabi-Yau of real dimension $10-D$.} In ten
dimensions, by requiring $\kappa$-symmetry of the GS action one finds
the constraints of ten-dimensional supergravity. In the non-critical
superstring, we get for backgrounds of this type one of the
supergravity constraints. In Appendix
\ref{sec:kappa-symm-and-torsion}, we work out the relation between
(\ref{eq:kappa-symmetry-condition}) and the torsion constraints.

\subsection{Classical integrability of the Green-Schwarz sigma-model}

In the following we construct a one-parameter family of flat currents
(\ref{flat}),(\ref{flatsolution}) that imply the existence of an
infinite number of conserved non-local charges, thus showing that the
GS sigma-model is classically integrable. The $\kappa$-invariance of
these currents will not be checked, but it should follow from the BRST
invariance of the corresponding pure-spinor currents shown in Appendix
\ref{sec:ps-charges-BRST-inv}.

The equation of motion and constraints for the currents that
follow from the action (\ref{eq:generic-GS-coset-action}) read
\begin{eqnarray}
  d{*J_2} & = & -J_0 \wedge *J_2 - *J_2 \wedge J_0 + J_1 \wedge J_1 -
  J_3 \wedge J_3 \ ,\\
  \lefteqn{J_1 \wedge J_2 + J_2 \wedge J_1 + *J_1 \wedge J_2 + J_2
  \wedge *J_1 = 0 \ ,}\\
  \lefteqn{J_2 \wedge J_3 + J_3 \wedge J_2 - J_2 \wedge *J_3 - *J_3
  \wedge J_2 = 0 \ ,} \ .
\end{eqnarray}

We are looking for a one parameter family of flat connections
$D=d+ a(\mu)$, satisfying the zero curvature condition $D^2=0$ or
in other words
 \be da(\mu)+a(\mu)\wedge a(\mu)=0,
 \ee
where the right-invariant current $a(\mu)$ is usually referred to
as the Lax connection and $\mu$ as the spectral parameter. In
order to facilitate the comparison with the pure spinor flat
current, we will switch to the left-invariant current $A = g^{-1}
a g$ which satisfies the equation
\begin{equation} \label{eq:GS-flat-current-eq}
  dA + A \wedge A + J \wedge A + A \wedge J = 0 \ .
\end{equation}
Following \cite{Bena:2003wd} we will consider a current composed
of the currents for which the exterior derivative is known:
\begin{equation}
  A = \alpha J_2 + \beta {*J_2} + \gamma J_1 + \delta J_3 \ .
\label{flat}
\end{equation}
Substituting this in (\ref{eq:GS-flat-current-eq}) and using the
equation of motion, the constraints and the Maurer-Cartan
equations yields the equations%
\footnote{Our currents are related to the currents in
  \cite{Bena:2003wd} by $p = -j_2$, $q = - (j_1 + j_3)$ and $q' = j_1
  - j_3$ so these equations are related to the ones in
  \cite{Bena:2003wd} by $\alpha = -\tilde \alpha$, $\beta = -\tilde
  \beta$, $\gamma = \tilde \delta - \tilde \gamma$ and $\delta =
  -(\tilde \gamma + \tilde \delta)$, where the tilded variables refer
  to the same untilded variables in \cite{Bena:2003wd}.}
\begin{eqnarray}
  \beta - \alpha + \gamma^2 + 2 \gamma & = & 0 \ ,\quad
  -\alpha - \beta + \delta^2 + 2 \delta = 0 \ , \nonumber\\
  -\gamma + (\alpha - \beta) \delta + \alpha - \beta + \delta & = & 0 \ ,
  \quad
  -\delta + (\alpha + \beta) \gamma + \alpha + \beta + \gamma = 0
  \ , \nonumber\\
  \alpha^2 - \beta^2 + 2 \alpha & = & 0 \ , \quad
  \gamma \delta + \gamma + \delta = 0 \ ,
\end{eqnarray}
whose two one-parameter families of solutions are
\begin{eqnarray}
  \alpha & = & 2 \sinh^2 \mu \ , \quad
  \beta = 2 \sinh \mu \cosh \mu \ , \quad
  \gamma = -(1 + e^{-\mu}) \ , \quad
  \delta = -(1 + e^\mu) \ , \nonumber \\
  \alpha & = & 2 \sinh^2 \mu \ , \quad
  \beta = -2 \sinh \mu \cosh \mu \ , \quad
  \gamma = e^\mu - 1 \ , \quad
  \delta = e^{-\mu} - 1 \ ,
\label{flatsolution}
\end{eqnarray}
where $-\infty < \mu < \infty$.

For the second family, an infinite set of conserved charges can be
obtained using the expansion of the solution about $\mu = 0$
\begin{equation}
  a = \mu (j_1 - j_3 - 2 {*j_2}) + \mu^2 \left( 2 j_2 +
  \frac{1}{2} j_1 + \frac{1}{2} j_3 \right) + O(\mu^3) \ ,
\end{equation}
where the $j_i$ denote the right-invariant currents $g J_i
g^{-1}$. We can then introduce the monodromy matrix, which is the
Wilson line of the flat connection
\begin{equation}
  U_C = \mathrm{P} \exp \left( -\int_C a \right) = 1 +
  \sum_{n=1}^\infty \mu^n Q_n \ ,
\end{equation}
whose expansion around $\mu=0$ leads to the conserved charges
$Q_n$. The first two conserved charges
are%
\footnote{In the notation of \cite{Bena:2003wd} the integrand of
$Q_1$ is proportial  the Noether current $p + \frac{1}{2} {*q'}$.}
\begin{eqnarray}
  Q_1 & = & -\int_C (j_1 - j_3 - 2 {*j_2}) \ , \\
  Q_2 & = & -\int_C \left( 2 j_2 + \frac{1}{2} j_1 + \frac{1}{2} j_3
  \right) + \nonumber\\
  && {} + \int_C [j_1 (x) -j_3 (x) - 2 {*j_2 (x)}] \int_o^x (j_1 -
  j_3 - 2 {*j_2}) \ .
\end{eqnarray}
The former is local and is expected to be one of the Noether
currents of the sigma-model. The latter is non-local. The other
charges can be generated by repetitive Poisson brackets of $Q_2$
and together they form a classical Yangian. The Lax connection is
the starting point for the solution of the classical sigma-model
(see e.g.\ \cite{Beisert:2005bm}).

We will not argue that the integrability property is preserved at
the quantum level. This will be shown in the pure spinor
formalism.

\section{Integrability of pure spinor superstrings on RR backgrounds}

In this section we will consider pure spinor superstrings on coset
super-manifolds $G/H$, where the supergroup $G$ possesses a
$\mathbb{Z}_4$ automorphism whose invariant locus is the subgroup
$H$. The cosets we will consider will be limited to backgrounds which
have only RR-flux. We will first discuss the various pure spinor
spaces in the different spacetime dimensions, then construct the BRST
invariant pure spinor action and the infinite set of BRST invariant
non-local charges, hence exhibiting the classical integrability of the
pure spinor superstrings. In the following section we will prove that
these pure spinor superstrings are also integrable at the quantum
level.  Towards the end of the section we will discuss the inclusion
of D-branes in the pure spinor superstrings.

The pure spinor formalism for the ten-dimensional superstring
\cite{Berkovits:2000fe} has been well established. In lower
dimensions, there have been different interpretations of the pure
spinor superstring action. In some cases it has been argued that
it describes the non-critical superstring \cite{Adam:2006bt}, in
other cases it has been argued to describe the non-compact sector
of a ten-dimensional superstring compactified on a CY manifold
\cite{Berkovits:2005bt,Grassi:2005sb,Wyllard:2005fh}. In this
section, we will focus on the algebraic properties of the pure
spinor formulation of the superstring of a supercoset sigma-model.

\subsection{Pure spinor spaces in two, four and six dimensions}
\label{puresection}

In this subsection we will present the definition of the pure
spinor spaces in lower-dimen\-sional superstrings. The definition of
the pure spinors that Cartan and Chevalley give in even dimension
$d=2n$ is that $\l\sigma^{m_1\ldots m_j}\l=0$ for $j<n$, so that
the pure spinor bilinear reads
\cite{cartan,chevalley}\footnote{See also
\cite{Berkovits:2004bw}.}
 \bea
 \l^\a\l^\b={1\over n!2^n} \sigma^{ \a\b}_{m_1\ldots
 m_n}(\l\sigma^{m_1\ldots m_n}\l),\label{cartan}
 \eea
where $\sigma^{m_1\ldots m_j}$ is the antisymmetrized product of
$j$ Pauli matrices. This definition of the pure spinor space in
$d=2,4,6$ dimensions is trivially realized by an $SO(d)$ Weyl
spinor.

In all our cases the lower-dimensional pure spinors will be Weyl
spinors. In some cases we will need more than just one pure spinor
to construct a consistent string theory. In particular, our pure
spinor spaces are dictated by the realization of the supersymmetry
algebra for the type II superstring.\footnote{These lower
dimensional pure spinors spaces have been introduced in
\cite{Berkovits:2005bt,Grassi:2005sb,Chandia:2005fi}, in the
context of the Calabi-Yau compactification of the ten-dimensional
pure spinor superstring.} Indeed, we will use the same pure spinor
spaces in $2p$ dimensions to describe the ghost sector of both the
non-compact sector of the type II superstring on $AdS_p\times
S^p\times CY_{5-p}$ and of the  $2p$ dimensional non-critical type
II superstring. These latter models have been introduced in
\cite{Adam:2006bt}, where a field redefinition has been
constructed that maps the RNS formulation to the pure spinor
formulation of the non-critical superstring in the linear dilaton
backgrounds. The crucial feature of these lower-dimensional pure
spinor spaces is that, like in the ten-dimensional case, the
product of two pure spinors is still proportional to the middle
dimensional form, according to \ref{cartan}. Let us discuss the
various dimensions in detail.

\subsubsection*{Two-dimensional superstring}

The left moving sector of Type II superstrings in two dimensions
realizes $\mathcal{N}=(2,0)$ spacetime supersymmetry with $2$ real supercharges
$Q_\a$, both of which are spacetime MW spinors of the same chirality,
which are related by an $SO(2)$ R--symmetry transformation ($\a$ is
not a spinor index in this case, but just enumerates supercharges of
the same chirality). The corresponding superderivatives are denoted by
$D_\a$. The supersymmetry algebra reads
 $$
 \{D_\a,D_\b\}=-\delta_{\a\b}P^+,
 $$ where $P^\pm$ are the holomorphic (antiholomorphic) spacetime
direction of $AdS_2$. The pure spinors are defined such that
$\l^\a D_\a$ is nilpotent, so that the pure spinor condition in
two dimensions reads \bea
\l^\a\l^\b\delta_{\a\b}=0,\label{2dpurity} \eea which is solved by
one Weyl spinor. The pure spinor bilinear reads \bea
\l^\a\l^\b&=&\half(\tau_a)^{\a\b}(\l^\g\tau^a_{\g\d}\l^\d), \eea
where the index $a$ takes the values $1, 3$. In two dimensions the
off-diagonal blocks of the gamma matrices are one dimensional
matrices, so the relation (\ref{cartan}) still holds.\footnote{The
notations here are slightly different from the ones in
\cite{Adam:2006bt}. In particular, if we denote by $\tilde\l^i$
the pure spinor in that paper, we have
$\tilde\l^1={1\over\sqrt{2}}(\l^1+i\l^2)$ and
$\tilde\l^2={1\over\sqrt{2}}(\l^1-i\l^2)$. Anyway, the pure spinor
space is identical to the one considered there.\label{ourfoot}}

\subsubsection*{Four-dimensional superstring}

In four dimensions, the left moving sector of the type II
superstring realizes ${\cal N}=1$ supersymmetry, which in terms of
the superderivatives $D_A$ in the Dirac form reads
 \bea
 \{D_A,D_B\}&=&-2(C\Gamma^m)P_m,
 \eea
where $C$ is the charge conjugation matrix and $A=1,\ldots,4$.
Requiring nilpotence of $\l^AD_A$ specifies the four-dimensional
pure spinor constraint
 \bea
 \l^A (C\Gamma^m)_{AB} \l^B&=&0.\label{4dpurity}
 \eea
If we expand the pure spinor bilinear in terms of the four
dimensional gamma matrices we find then
 $\lambda^A\lambda^B={1\over4}(C\Gamma_{mn})^{AB}(\l C\Gamma^{mn}\l)$.
Sometimes it will be convenient to use the Weyl notation for the
spinors, under which the pure spinor is represented by a pair of
Weyl and anti-Weyl spinors $(\l^\a,\l^\ad)$, subject to the
constraint
 \bea
 \l^\a\l^\ad=0.
 \eea
The pure spinor bilinear then reads
 \bea
 \l^\a\l^\b={1\over 8}\sigma_{mn}^{\a\b}(\l\sigma^{mn}\l),&\quad&
 \l^\ad\l^\bd={1\over 8}\sigma_{mn}^{\ad\bd}(\l\sigma^{mn}\l),
 \eea

\subsubsection*{Six-dimensional superstring}

In six dimensions, the left moving sector of the type II
superstring realizes ${\cal N}=(1,0)$ supersymmetry, with eight
real supercharges. Naively, one would expect that one supercharge
$Q_\a$ in the ${\bf \bar 4}$ of $SO(6)$ could do the job. However,
due to CPT invariance and the pseudo-reality of the Weyl irrep, it
is impossible to realizes the supersymmetry algebra with just one
copy of supercharges\footnote{A simple manifestation of this fact
is the following. The six-dimensional Pauli matrices
$\sigma^m_{\a\b}$ are four by four antisymmetric matrices.
Therefore the naive supersymmetry algebra
$\{Q_\a,Q_\b\}=\sigma^m_{\a\b}P_m$ does not make sense in six
dimensions.} and we have to introduce two supercharges $Q^i_\a$ in
the ${\bf \bar 4}$ of $SO(6)$, which form a doublet of an
auxiliary $SU(2)$ outer automorphism. In terms of the
superderivatives $D_\a^i$ the supersymmetry algebra reads
 \bea
 \{D_\a^i,D_\b^j\}=\e^{ij}\sigma^m_{\a\b}P_m,
 \eea
where $\e^{ij}$ is the invariant tensor of $SU(2)$. It is clear
now that the six-dimensional pure spinor consists of a Weyl spinor
$\l^\a_i$ which is also a doublet with respect to the auxiliary
$SU(2)$. If we demand the nilpotence of $\l^\a_i D^i_\a$ we then
find the pure spinor constraint
 \bea
 \e^{ij}\l^\a_i\sigma^m_{\a\b}\l^\b_j=0.\label{purity6}
 \eea
If we expand the symmetric bispinor constructed out of a pure
spinor bilinear, using representation theory we find
once again that only the middle-dimensional form is present
 \bea
 \l^\a_i\l^\b_j&=&{1\over
 3!16}\sigma_{mnp}^{\a\b}\sigma^{ab}_{ij}(\l\sigma^{mnp}\sigma_{ab}\l),
 \eea
where $\sigma^{ab}_{ij}$ is the two by two $SU(2)$ generator in
the fundamental representation, given by the antisymmetrized
product of two $SU(2)$ Pauli matrices.

\subsection{The pure spinor sigma-model} \label{sec:pure-spinor-sigma-model}

The worldsheet action in the pure spinor formulation of the
superstring consists of a matter and a ghost sector. The
worldsheet metric is in the conformal gauge and there are no
reparameterization ghosts. The matter fields are written in terms
of the left-invariant currents $J = g^{-1}
\partial g$, $\bar J = g^{-1} \bar \partial g$, where $g: \Sigma
\to G$, and decomposed according to the invariant spaces of the
$\mathbb{Z}_4$ automorphism:
\begin{equation}
  J = J_0 + J_1 + J_2 + J_3
\end{equation}
and similarly for the anti-holomorphic component $\bar J$, where
the notations are the same as in section 2. The Lie algebra-valued
pure spinor fields and their conjugate momenta are defined as in
\cite{Berkovits:2004xu}
\begin{equation}
  \lambda = \lambda^\alpha T_\alpha \ ,\quad
  w = w_\alpha \eta^{\alpha \hat \alpha} T_{\hat \alpha} \ , \quad
  \bar \lambda = \bar \lambda^{\hat \alpha} T_{\hat \alpha} \ , \quad
  \bar w = \bar w_{\hat \alpha} \eta^{\alpha \hat \alpha} T_\alpha \
  ,\label{purepara}
\end{equation}
where we decomposed the fermionic generators $T$ of the super Lie
algebra $\mathcal{G}$ according to their $\mathbb{Z}_4$ gradings
$T_\alpha \in \mathcal{H}_1$ and $T_{\hat \alpha} \in \mathcal{H}_3$
and used the inverse of the Cartan metric $\eta^{\alpha \hat
\alpha}$. The spinor indices here are just a reminder, the unhatted
ones refer to left moving quantities, the hatted ones to right moving
ones.  The choice of spinor representations depends on the particular
supercoset in discussion and will be explained in Section
\ref{examples} for each specific model. Using these conventions, the
pure spinor currents are defined by
\begin{equation}
  N = - \{ w, \lambda \} \ , \quad \bar N = - \{ \bar w, \bar \lambda
  \} \ ,
\end{equation}
which generate in the pure spinor variables the Lorentz
transformations that correspond to left-multiplication by elements
of $H$. $N, \bar N \in \mathcal{H}_0$ so they indeed act on the
tangent-space indices $\alpha$ and $\hat \alpha$ of the pure
spinor variables as the Lorentz transformation. The pure spinor
constraint reads
 \be
 \{\l,\l\}=0,\qquad \{\bar\l,\bar\l\}=0 \ .
 \ee

The sigma-model should be invariant under the global
transformation $\delta g = \Sigma g$, $\Sigma \in \mathcal{G}$.
$J$ and $\bar J$ are invariant under this global symmetry. The
sigma-model should also be invariant under the gauge
transformation
\begin{eqnarray}
  \delta_\Lambda J & = & \partial \Lambda + [J, \Lambda] \ ,\quad
  \delta_\Lambda \bar J = \bar \partial \Lambda + [\bar J, \Lambda]
  \,\quad
  \delta_\Lambda \lambda = [\lambda, \Lambda] \ , \quad
  \delta_\Lambda w = [w, \Lambda] \ , \nonumber\\
  \delta_\Lambda \bar \lambda & = & [\bar \lambda, \Lambda] \ ,
  \quad
  \delta_\Lambda \bar w = [\bar w, \Lambda] \ ,
\end{eqnarray}
where $\Lambda \in \mathcal{H}_0$. The most general sigma-model
with these properties is
\begin{equation}
  S = \int d^2 z \langle \alpha J_2 \bar J_2 + \beta J_1 \bar J_3 +
  \gamma J_3 \bar J_1 + w \bar \partial \lambda + \bar w \partial \bar
  \lambda + N \bar J_0 + \bar N J_0 + a N \bar N \rangle\,,
\end{equation}
where $\alpha,\beta,\gamma,a$ are numerical coefficients that we
will shortly determine.

The accompanying BRST operator is (see
Appendix~\ref{sec:pure-spinor-sigma-model-from-BRST})
\begin{equation}
  Q_B = \oint \langle dz \lambda J_3 + d\bar z \bar \lambda \bar J_1
  \rangle \ ,
\end{equation}
which generates the following BRST transformations
 \bea
\label{brstrans}\delta_B J_j & =& \delta_{j+3,0}\partial
(\epsilon\l)+[J_{j+3},\epsilon\l]+\delta_{j+1,0}\partial(\epsilon\bar\l)+[J_{j+1},\epsilon\bar\l],\\
\delta_B \bar J_j & = & \delta_{j+3,0}\bar \partial (\epsilon\l)+[\bar
J_{j+3},\epsilon\l]+\delta_{j+1,0}\bar \partial(\epsilon\bar\l)+[\bar
J_{j+1},\epsilon\bar\l],\nonumber\\ \delta_B w& = & -J_3\epsilon,
\qquad \delta_B \bar w=-\bar J_1\epsilon,\nonumber\\ \delta_B N & = &
[J_3,\epsilon\l],\qquad \delta_B \bar N=[\bar J_1,\epsilon\bar
\l].\nonumber \eea The coefficients of the various terms in the action
are determined by requiring that the action be BRST invariant (the
details can be found in
Appendix~\ref{sec:pure-spinor-sigma-model-from-BRST}). The
BRST-invariant sigma-model thus obtained is
\begin{equation} \label{eq:AdS-ps-action}
  S = \int d^2 z \left\langle \frac{1}{2} J_2 \bar J_2 + \frac{1}{4} J_1
  \bar J_3 + \frac{3}{4} J_3 \bar J_1 + w \bar\partial  \lambda + \bar w
  \partial \bar \lambda + N \bar J_0 + \bar N J_0 -N \bar N
  \right\rangle
\end{equation}
for all dimensions and this of course matches the critical $AdS_5
\times S^5$ considered in \cite{Berkovits:2004xu} as well.

Let us briefly comment on the relation between the pure spinor
action (\ref{eq:AdS-ps-action}) and the GS action
(\ref{eq:generic-GS-coset-action}). The latter, when written in
conformal gauge, reads
\begin{equation}
  S_\mathrm{GS} = \int d^2 z \langle \frac{1}{2} J_2 \bar J_2 +
  \frac{1}{4} J_1 \bar J_3 - \frac{1}{4} J_3 \bar J_1 \rangle \ .
\end{equation}
To this one has to add a term which breaks $\kappa$-symmetry and
adds kinetic terms for the target-space fermions and coupling to
the RR-flux $P^{\a\ah}$
\begin{equation}
  S_\kappa = \int d^2 z (d_\alpha \bar J_1^\alpha + \bar d_{\hat \alpha}
  J_3^{\hat \alpha} + P^{\alpha \hat \alpha} d_\alpha \bar d_{\hat
  \alpha}) = \int d^2 z \langle d \bar J_1 - \bar d J_3 + d \bar d
  \rangle \ ,
\end{equation}
where, in curved backgrounds, the $d$'s are the conjugate
variables to the superspace coordinates $\theta$'s. After
integrating out $d$ and $\bar d$ we get the complete matter part
\begin{equation}
  S_\mathrm{GS} + S_\kappa = \int d^2 z \langle \frac{1}{2} J_2 \bar
  J_2 + \frac{1}{4} J_1 \bar J_3 + \frac{3}{4} J_3 \bar J_1 \rangle \ .
\end{equation}
This sigma-model can be recognized as taking the same form as the
sigma-model used in \cite{Berkovits:1999zq} for the
compactification of type II superstring on $AdS_2 \times S^2\times
CY_3$ in the hybrid formalism. It is a general fact that the
matter part of the hybrid and the pure spinor formalism is the
same. As usual this has to be supplemented with kinetic terms for
the pure spinors and their coupling to the background
 \begin{equation}
 S_{gh}=\int d^2 z \left\langle  w \bar\partial  \lambda + \bar w
  \partial \bar \lambda + N \bar J_0 + \bar N J_0 -N \bar N
  \right\rangle
\end{equation}
in order to obtain the full superstring sigma-model
(\ref{eq:AdS-ps-action}) with action $S=S_{GS}+S_\kappa+S_{gh}$.

\subsection{Classical integrability of the pure spinor sigma-model}

In this subsection we will demonstrate the classical integrability
of the action (\ref{eq:AdS-ps-action}).  For finding the equations
of motion and the flat currents we follow the method of
\cite{Vallilo:2003nx}. Here, one has to distinguish between two
cases --- a non-Abelian gauge symmetry $H$ and an Abelian one,
which occurs only in the two-dimensional non-critical
superstrings. We begin with the non-Abelian case and then discuss
the differences when the gauge group is Abelian.

The equations of motion of the currents $J_i$ are obtained by
considering the variation $\delta g = g X$ under which $\delta J =
\partial X + [J, X]$ and using the $\mathbb{Z}_4$ grading and the
Maurer-Cartan equations, so that we get
\begin{eqnarray}
  \nabla \bar J_3 & = & - [J_1, \bar J_2] - [J_2, \bar J_1] + [N, \bar
  J_3] + [\bar N, J_3] \ ,\\
  \bar \nabla J_3 & = & [N, \bar J_3] + [\bar N, J_3] \ ,\\
  \nabla \bar J_2 & = & - [J_1, \bar J_1] + [N, \bar J_2] + [\bar N,
  J_2] \ ,\\
  \bar \nabla J_2 & = & [J_3, \bar J_3] + [N, \bar J_2] + [\bar N,
  J_2] \ ,\\
  \nabla \bar J_1 & = & [N, \bar J_1] + [\bar N, J_1] \ ,\\
  \bar \nabla J_1 & = & [J_2, \bar J_3] + [J_3, \bar J_2] + [N, \bar
  J_1] + [\bar N, J_1] \ ,
\end{eqnarray}
where $\nabla J = \partial J + [J_0, J]$ and $\bar \nabla J = \bar
\partial J + [\bar J_0, J]$ are the gauge covariant derivatives.
The equations of motion of the pure spinors and the pure spinor
gauge currents are
\begin{eqnarray}
  \bar \nabla \lambda & = & [\bar N, \lambda] \ , \quad
  \nabla \bar \lambda = [N, \bar \lambda] \ ,\\
  \bar \nabla N & = & -[N, \bar N] \ , \quad \nabla \bar N = [N, \bar
  N]  \ . \label{eq:AdS-N-eq}
\end{eqnarray}

As in the previous section on the GS formalism, we are looking for
a one-parameter family of right-invariant flat currents $a(\mu)$.
The left-invariant current $A = g^{-1} a g$ constructed from the
flat current $a$ satisfies the equation
\begin{equation} \label{eq:AdS-ps-flatness}
  \nabla \bar A - \bar \nabla A + [A, \bar A] +
  \sum_{i=1}^3 \left( [J_i, \bar A] + [A, \bar J_i] \right) = 0 \ .
\end{equation}
$A$ and $\bar A$ can depend on all the currents for which there
are equations of motion so
\begin{equation}
  A = c_2 J_2 + c_1 J_1 + c_3 J_3 + c_N N \ , \quad
  \bar A = \bar c_2 \bar J_2 + \bar c_1 \bar J_1 + \bar c_3 \bar J_3 + \bar c_N \bar N \ .
\end{equation}
By requiring the coefficients of the currents to satisfy
(\ref{eq:AdS-ps-flatness}) one obtains the equations
\begin{eqnarray}
  && - \bar c_2 + c_1 \bar c_1 + \bar c_1 + c_1 = 0 \ , \quad
  -\bar c_3 + c_1 \bar c_2 + \bar c_2 + c_1 = 0 \, \quad
  -\bar c_3 + c_2 \bar c_1 + \bar c_1 + c_2 = 0 \ , \nonumber \\
  && - c_2 + c_3 \bar c_3 + \bar c_3 + c_3 = 0 \, \quad
  -c_1 + c_2 \bar c_3 + \bar c_3 + c_2 = 0 \ , \quad
  -c_1 + c_3 \bar c_2 + \bar c_2 + c_3 = 0 \ ,\nonumber \\
  && \bar c_1 - c_1 + c_N \bar c_1 + c_N = 0 \ , \quad
  \bar c_1 - c_1 - c_1 \bar c_N - \bar c_N = 0 \ , \quad
  \bar c_2 - c_2 + c_N \bar c_2 +c_N = 0 \ ,\nonumber \\
  && \bar c_2 - c_2 - c_2 \bar c_N - \bar c_N =0 \, \quad
  \bar c_3 - c_3 + c_N \bar c_3 + c_N = 0 \ , \quad
  \bar c_3 - c_3 - c_3 \bar c_N -\bar c_N = 0 \ ,\nonumber \\
  && c_2 \bar c_2 + \bar c_2 + c_2 = 0 \ , \quad
  c_1 \bar c_3 + \bar c_3 + c_1 = 0 \ , \quad
  c_3 \bar c_1 + \bar c_1 + c_3 = 0 \ , \nonumber \\
  && \bar c_N + c_N + c_N \bar c_N = 0 \ , \label{eq:AdS-ps-flat-current-eqns}
\end{eqnarray}
whose solutions can be written as
\begin{eqnarray}
  c_2 & = & \mu^{-1} - 1 \ , \quad
  c_1 = \pm \mu^{-1/2} - 1 \ , \quad
  c_3 = \pm \mu^{-3/2} - 1 \ , \quad
  \bar c_2 = \mu - 1 \ , \nonumber \\
  \bar c_1 & = & \pm \mu^{3/2} - 1 \ , \quad
  \bar c_3 = \pm \mu^{1/2} - 1 \ , \quad
  c_N = \mu^{-2} - 1 \ , \quad
  \bar c_N = \mu^2 - 1 \ . \label{eq:AdS-flat-current-params}
\end{eqnarray}
Hence, there exists a one-parameter set of flat currents.

The flat currents are given by the right-invariant versions $a = g
A g^{-1}$ and $\bar{a} = g \bar A g^{-1}$ of the currents $A$ and
$\bar A$ found above. The conserved charges are given by
\begin{equation} \label{eq:ps-conserved-charges}
  U_C = \mathrm{P} \exp \left[ -\int_C \left( dz a + d\bar z
  \bar{a} \right) \right] \ .
\end{equation}
These charges should be BRST-closed in order to represent physical
symmetries. In Appendix \ref{sec:ps-charges-BRST-inv} it is shown
that these charges are indeed BRST invariant.

The construction of the flat currents in the case of an Abelian
gauge group is very similar with some differences we will now
discuss. The equations of the pure spinor gauge generators
(\ref{eq:AdS-N-eq}) degenerate in the Abelian case into the
equations
\begin{equation}
  \bar \partial N = 0 \ , \quad \partial \bar N = 0 \ .
\end{equation}
As a result, the last equation in
(\ref{eq:AdS-ps-flat-current-eqns}) drops. However, the solution
(\ref{eq:AdS-flat-current-params}) remains valid. The proof of the
classical BRST invariance of these charges is identical to the one
in the non-Abelian case.

The first two conserved charges can be obtained by expanding $\mu
= 1 + \epsilon$ about $\epsilon = 0$. To simplify the notation we
will consider the right invariant currents
\begin{equation}
  j_i \equiv g J_i g^{-1} \ , \quad
  \bar j_i \equiv g \bar J_i g^{-1} \ , \quad
  n \equiv g N g^{-1} \ , \quad
  \bar n \equiv g \bar N g^{-1} \ .
\end{equation}
Using the expansion in $\epsilon$ one gets
\begin{eqnarray}
  a & = & -\left( \frac{1}{2} j_1 + j_2 + \frac{3}{2} j_3 + 2 n
  \right) \epsilon + \left( \frac{3}{8} j_1 + j_2 + \frac{15}{8} j_3 + 3 n
  \right) \epsilon^2 + O(\epsilon^3) \ , \\
  \bar {a} & = & \left( \frac{3}{2} \bar j_1 + \bar j_2 +
  \frac{1}{2} \bar j_3 + 2 \bar n \right) \epsilon + \left(
  \frac{3}{8} \bar j_1 - \frac{1}{8} \bar j_3 + \bar n \right)
  \epsilon^2 + O(\epsilon^3) \ ,
\end{eqnarray}
whose substitution in (\ref{eq:ps-conserved-charges}) and using
$U_C = 1 + \sum_{n = 1}^\infty \epsilon^n Q_n$ yields
\begin{eqnarray}
  Q_1 & = & \int_C \left[ dz \left( \frac{1}{2} j_1 + j_2 +
  \frac{3}{2} j_3 + 2 n \right) - d\bar z \left( \frac{3}{2} \bar j_1
  + \bar j_2 + \frac{1}{2} \bar j_3 + 2 \bar n \right) \right] \ ,\\
  Q_2 & = & - \int_C \left[ dz \left( \frac{3}{8} j_1 + j_2 +
  \frac{15}{8} j_3 + 3 n \right) + d\bar z \left( \frac{3}{8} \bar j_1
  - \frac{1}{8} \bar j_3 + \bar n \right) \right] + \nonumber \\
  && {} + \int_C \left[ dz \left( \frac{1}{2} j_1 + j_2 + \frac{3}{2} j_3 +
  2 n \right) \bigg|_{(z, \bar z)} - d \bar z \left( \frac{3}{2} \bar
  j_1 + \bar j_2 + \frac{1}{2} \bar j_3 + 2 \bar n \right) \bigg|_{(z,
  \bar z)} \right] \times \nonumber \\
  && {} \times \int_o^{(z, \bar z)} \Bigg[ dz' \left( \frac{1}{2} j_1
  + j_2 + \frac{3}{2} j_3 + 2 n \right) \bigg|_{(z', \bar z')} -
  \nonumber\\
  && {} - d\bar z' \left( \frac{3}{2} \bar j_1 + \bar j_2 +
  \frac{1}{2} \bar j_3 + 2 \bar n \right) \bigg|_{(z', \bar z')}
  \Bigg] \ .
\end{eqnarray}
The first charge $Q_1$ is the local Noether charge. The rest of
the conserved charges, which form the Yangian algebra, can be
obtained by repetitive commutators of $Q_2$.

\subsection{Adding D-branes to the pure spinor superstrings}
\label{secD}

In this section we will consider the addition of D-branes to the
coset space background. For this purpose we consider the
implications of adding boundaries to the worldsheet (adding
D-branes in the pure spinor formalism is treated in
\cite{Berkovits:2002ag,Schiappa:2005mk}) and requiring the
appropriate boundary conditions.

The contribution of the boundary to the variation $\delta g = g X$
of the pure spinor action is
\begin{eqnarray}
  \delta S & = & i \oint_{\partial \Sigma} \Big\langle (\frac{1}{4} d \bar z
  \bar J_3 - \frac{3}{4} dz J_3) X_1 + \frac{1}{2} (d \bar z \bar J_2
  - dz J_2) X_2 +(\frac{3}{4} d \bar z \bar J_1 - \frac{1}{4} dz J_1)
  X_3 + \nonumber \\
  && {} + d \bar z \bar w \delta \bar \lambda - dz w \delta \lambda
  \Big\rangle + \dots \ ,
\end{eqnarray}
where $X$ has been decomposed into its $\mathbb{Z}_4$ invariant
components $X_i$ and the $\dots$ are the worldsheet bulk terms. We
will consider the worldsheet as the upper-half complex plane so
the boundary is given by $z = \bar z$. The boundary conditions
that follow are
\begin{equation} \label{eq:ps-boundary-eq}
  \bar J_3 \Big|_{\partial \Sigma} = 3 J_3 \Big|_{\partial \Sigma} \ ,
  \quad
  J_2 \Big|_{\partial \Sigma} = \bar J_2 \Big|_{\partial \Sigma} \ ,
  \quad
  J_1 \Big|_{\partial \Sigma} = 3 \bar J_1 \Big|_{\partial \Sigma} \ ,
  \quad
  w_\alpha \delta \lambda^\alpha \Big|_{\partial \Sigma} = -\bar
  w_{\hat \alpha} \delta \bar \lambda^{\hat \alpha} \Big|_{\partial \Sigma}
  \ .
\end{equation}
An additional constraint comes from requiring the action to be
BRST invariant. The BRST variation of the action is
\begin{equation}
  \delta_B S = \frac{i}{4} \oint_{\partial \Sigma} \langle d \bar z
  \epsilon (\lambda \bar J_3 - \bar \lambda \bar J_1) - dz \epsilon (
  \bar \lambda J_1 - \lambda J_3) \rangle \ ,
\end{equation}
which after substituting (\ref{eq:ps-boundary-eq}) takes the form
\begin{equation}
  \delta_B S = i \oint_{\partial \Sigma} ( dz \epsilon j_B - d \bar z
  \epsilon \bar j_B) \ ,
\end{equation}
so we have to require in addition $j_B = \bar j_B$ on the
boundary.

We may solve the pure spinor boundary conditions by
\begin{equation}
  (\lambda^\alpha - R^\alpha{}_{\hat \alpha} \bar \lambda^{\hat
  \alpha}) \Big|_{\partial \Sigma} = 0 \ , \quad
  (w_\alpha + R_\alpha{}^{\hat \alpha} \bar w_{\hat \alpha})
  \Big|_{\partial \Sigma} = 0 \ ,
\end{equation}
in which the matrix $R_\alpha{}^{\hat \alpha}$ determines the type
of D-brane and $R_\alpha{}^{\hat \alpha} R^\alpha{}_{\hat \beta} =
\delta^{\hat \alpha}_{\hat \beta}$. The BRST boundary condition
then becomes
\begin{equation}
  (\bar J_1^\alpha - R^\beta{}_{\hat \beta} \eta_{\beta \hat \alpha}
  \eta^{\alpha \hat \beta} J_3^{\hat \alpha}) \Big|_{\partial \Sigma}
  = 0 \ .
\end{equation}
This condition can also be obtained by requiring that the boundary
condition involving $w$ and $\bar w$ be BRST invariant.

The matrix $R$ is to be determined by the symmetries that the
D-brane configuration breaks. However, since the boundary
conditions for the matter fields (\ref{eq:ps-boundary-eq}) involve
only left-invariant currents, they alone are not sufficient in
order to break some of the symmetries. In order to gain such
information it is probably necessary to resort to a specific
parameterization of the super-Lie manifold $G$ and the gauged
subgroup $H$.

\section{Quantum consistency of the pure spinor sigma-model}

In this section we will show that the pure spinor superstring on
the supercoset backgrounds in various dimensions is gauge
invariant and BRST invariant to all orders in the sigma-model
perturbation theory. Then, we will show that the infinite set of
nonlocal charges, which are classically conserved, are also BRST
invariant in the quantum theory, proving that the integrability of
the superstring holds quantum mechanically as well.

Since our backgrounds are realized in terms of supercosets with a
$\ZZ_4$ automorphism, we will be able to apply the powerful tools
developed in \cite{Berkovits:2004xu} for the superstring on the
$AdS_5\times S^5$ background. The only subtlety is related to the
different definitions of the pure spinor constraints in the lower
dimensional cases.

\subsection{Quantum gauge invariance}

As we discussed above, the action is classically gauge invariant
under the right multiplication $g\to gh$, where $h\in H$. We will
prove that we can always add a local counterterm such that the
quantum effective action remains gauge invariant at the quantum
level.\footnote{This proof is different from the one in
\cite{Berkovits:2004xu}. In that paper, Berkovits uses a parity
symmetry argument, while we use the consistency condition on gauge
anomalies.} Quantum gauge invariance will then be used to prove
BRST invariance.

An anomaly in the $H$ gauge invariance would show up as a
nonvanishing gauge variation of the effective action
$\delta_\Lambda S_{eff}$ in the form of a local operator. Since
there is no anomaly in the global $H$ invariance, the variation
must vanish when the gauge parameter is constant and, moreover, it
must have grading zero. Looking at the list of our worldsheet
operators, we find that the most general form of the variation is
 \bea
 \delta S_{eff}&=&\int d^2z\langle c_1 N\bar\partial \Lambda+\bar c_1\bar
 N\partial \Lambda+2c_2J_0\bar\partial\Lambda+2\bar c_2\bar
 J_0\partial\Lambda\rangle,\label{lambdas}
 \eea
where $\Lambda=T_{[ab]}\Lambda^{[ab]}(z,\bar z)$ is the local
gauge parameter and $(c_1,\bar c_1,c_2,\bar c_2)$ are arbitrary
coefficients. By adding the counterterm
 \bea
 S_c&=&-\int d^2z\langle c_1 N\bar J_0+\bar c_1\bar NJ_0+(c_2+\bar
 c_2)J_0\bar J_0\rangle,
 \eea
we find that the total variation becomes
 \bea
 \delta_\Lambda (S_{eff}+S_c)&=&(c_2-\bar c_2)\int d^2z\langle
 J_0\bar\partial\Lambda-\bar J_0\partial\Lambda\rangle.
 \eea
On the other hand, the consistency condition on the gauge anomaly
requires that
 \bea
( \delta_\Lambda\delta_{\Lambda'}-\delta_{\Lambda'}\delta_\Lambda)
S_{eff}=\delta_{[\Lambda,\Lambda']}S_{eff},
 \eea
which fixes the coefficients $c_2=\bar c_2$.
Therefore the action is gauge invariant quantum mechanically.

\subsection{Quantum BRST invariance}
\label{quantumbrst}

In order to prove the BRST invariance of the superstring at all
orders in perturbation theory we will adapt the proof of
\cite{Berkovits:2004xu} to our lower-dimensional cases. First, we
will show that the classical BRST charge is nilpotent. We will
then prove that the effective action can be made classically BRST
invariant by adding a local counterterm, using triviality of a
classical cohomology class. Then we will prove that order by order
in perturbation theory no anomaly in the BRST invariance can
appear.

As we have shown in the previous section, the action
(\ref{eq:AdS-ps-action}) in the pure spinor formalism is
classically BRST invariant. It is easy to prove, following the
algebraic argument \cite{Berkovits:2004xu}, that, in all our
backgrounds, the pure spinor BRST charge is classically nilpotent
on the pure spinor constraint, up to gauge invariance and the
ghost equations of motion. The second variation of the ghost
currents reads indeed
 \bea
 Q^2(N)&=&-[N,\Lambda]-\{\l,\nabla\bar\l-[N,\bar\l]\},\nn\\
 Q^2(\bar N)&=&-[\bar N,\Lambda]-\{\bar \l,\bar\nabla\l-[\bar N,\l]\},
 \eea
for the particular gauge transformation parameterized by
$\Lambda=\{\l,\bar\l\}$ and the equations of motion
(\ref{eq:AdS-N-eq}). Therefore the classical BRST charge is well defined.

Consider now the quantum effective action $S_{eff}$. After the
addition of a suitable counterterm, it is gauge invariant to all
orders. Moreover, the classical BRST transformations of
(\ref{brstrans}) commute with the gauge transformations, since the
BRST charge is gauge invariant. Therefore, the anomaly in the
variation of the effective action, which is a local operator, must
be a gauge invariant integrated vertex operator of ghost number
one
 \bea
  \delta_{BRST}S_{eff}&=&\int d^2z\langle {\Omega}_{z\bar z}^{(1)}\rangle.
  \eea
In Appendix \ref{empti} we show that the cohomology of such
operators is empty, namely that we can add a local counterterm to
cancel the BRST variation of the action. A crucial step in the
proof is that the symmetric bispinor, constructed with the product
of two pure spinors, is proportional to the middle dimensional
form. Schematically, this means that in $d=2n$ dimensions we can
decompose
$$ \l^\a\l^\b\sim\sigma_{m_1\ldots
m_n}^{\a\b}(\l\sigma^{m_1\ldots m_n}\l).
 $$
In section \ref{puresection}, we have shown that this property is
satisfied by the pure spinors in all our backgrounds, ensuring
classical BRST invariance of the effective action.

Since there are no conserved currents of ghost number two in the
cohomology that could deform $Q^2$ the quantum modifications to
the BRST charge can be chosen such that its nilpotence is
preserved. In this case, we can set the anti-fields to zero and
use algebraic methods to extend the BRST invariance of the
effective action by induction to all orders in perturbation
theory. Suppose the effective action is invariant to order
$h^{n-1}$. This means that
 $$\tilde Q S_{eff}=h^n\int d^2z\langle {\Omega}^{(1)}_{z\bar
z}\rangle+{\cal O}(h^{n+1}).
$$
The quantum modified BRST operator $\tilde Q=Q+Q_q$ is still
nilpotent up to the equations of motion and the gauge invariance.
This implies that $Q\int\,d^2z\langle\Omega^{(1)}_{z\bar
z}\rangle=0$. But the cohomology of ghost number one integrated
vertex operators is empty, so $\Omega^{(1)}_{z\bar z}=Q
\Sigma^{(0)}_{z\bar z}$, which implies
 \bea
 \tilde Q\left(S_{eff}- h^n\int\,d^2z\langle\Sigma^{(0)}_{z\bar
 z}\rangle\right)&=& {\cal O}(h^{n+1}).
 \eea
Therefore, order by order in perturbation theory it is possible to
add a counterterm that restores BRST invariance.

\subsection{Quantum integrability}

In this subsection we will finally show that the classically
conserved nonlocal currents of (\ref{eq:AdS-flat-current-params})
can be made BRST invariant quantum mechanically. In this way we
prove quantum integrability of our type II superstring theories.
The proof is essentially identical to the one presented in
\cite{Berkovits:2004xu}. First, we review how the absence of a
certain ghost number two state from the cohomology implies the
existence of an infinite number of nonlocal BRST invariant
charges. Then we will review how this argument can be extended
quantum mechanically.

Consider the charge that generates the global symmetry with
respect to the supergroup $G$
 \bea
 q\equiv q^AT_A=\int d\sigma j^A T_A,
 \eea
where $j^A$ is the corresponding gauge invariant current. Since
this is a symmetry of the theory, the charge is BRST invariant, so
we find $\e Qj=\partial_\sigma h$, where $h=h^A T_A$ is a certain
operator of ghost number one and weight zero. Classical nilpotence
of the BRST charge implies moreover that $Q h=0$.

Consider now the operator $:\{h,h\}:$, where $:\ldots:$ denotes a
BRST invariant normal ordering prescription. If there exists a
ghost number one and weight zero operator $\Omega$, such that
 \bea
 Q\Omega=:\{h,h\}:,\label{omeh}
 \eea
then there is an infinite number of nonlocal charges which are
classically BRST invariant. To prove this, consider the nonlocal
operator
 \bea
 k=:\int_{-\infty}^{+\infty}d\sigma\int_{-\infty}^\sigma
 d\sigma'[j(\sigma),j(\sigma')]:.
 \eea
 Its BRST variation is
 $Qk=2:\int_{-\infty}^{+\infty}d\sigma[j(\sigma),h(\sigma)]:$. On
 the other hand, the BRST transformations are classically
 nilpotent, in fact we find
 $Q(2:[j(\sigma),h(\sigma)]:)=\partial_\sigma:\{h(\sigma),h(\sigma)\}:$.
Now, since there is an operator $\Omega$ that satisfies
(\ref{omeh}), we have
 \bea
 Q(2:[j,h]:-\partial_\sigma\Omega)=0.
 \eea
In other words, the ghost number one weight one operator
$2:[j,h]:-\partial_\sigma\Omega$ is BRST closed. On the other
hand, the BRST cohomology of ghost number one currents ${\cal
O}^{(1)}_\sigma$ is empty, as we will show below. We conclude that
this operator is BRST exact, namely there exists a $\Sigma^{(0)}$
such that $Q\Sigma^{(0)}=2:[j,h]:-\partial_\sigma\Omega$. But then
the nonlocal charge
 \bea
 \tilde q&=&k-\int_{-\infty}^{+\infty} d\sigma\Sigma,
 \eea
is classically BRST invariant and represent the first nonlocal
charge of the Yangian. By commuting $\tilde q$ with itself one
generates the whole Yangian.

It remains to be shown that the BRST cohomology of ghost number
one currents is trivial. This cohomology, in fact, is equivalent
to the cohomology of ghost number two unintegrated vertex
operators, by the usual descent relation
 \bea
 Q \int d\sigma {\cal
O}^{(1)}_\sigma=0 \Rightarrow Q {\cal
O}^{(1)}_\sigma=\partial_\sigma {\cal O}^{(2)}.
 \eea
At ghost number two we have only two unintegrated vertex operators
that transform in the adjoint of the global supergroup $G$, namely
 \bea
 V_1=g\l\bar\l g^{-1},&\quad& V_2=g\bar \l\l g^{-1}.
 \eea
 Their sum is BRST closed, while their difference is not. Finally,
 we have $ V_1+V_2=Q \Omega^{(1)}$ where
 \bea
 \Omega^{(1)}&=&\half g(\l+\bar\l)g^{-1},
 \eea
 so this classical cohomology class is empty.

Now, suppose that we have a BRST invariant nonlocal charge $q$ at
order $h^{n-1}$ in perturbation theory, namely $\tilde Q
q=h^n\Omega^{(1)}+{\cal O}(h^{n+1})$. $\Omega^{(1)}$ must be a
ghost number one local charge, since any anomaly must be
proportional to a local operator. Nilpotence of the quantum BRST
charge $\tilde Q=Q+Q_q$ implies that $Q\Omega^{(1)}=0$, but the
classical cohomology at ghost number one and weight one is empty,
as shown above, so there exists a current $\Sigma^{(0)}(\sigma)$
such that $Q\int d\sigma\Sigma^{(0)}(\sigma)=\Omega^{(1)}$. As a
result $\tilde Q(q-h^n\int d\sigma\Sigma^{(0)}(\sigma))={\cal
O}(h^{n+1})$. Hence, we have shown that it is possible to modify
the classically BRST invariant charges of
(\ref{eq:AdS-flat-current-params}) such that they remain BRST
invariant at all orders in perturbation theory.

\section{One-loop conformal invariance} \label{sec:one-loop-conformal-inv}

In this section we will give the spacetime interpretation of the
various sigma-models we have introduced in the previous sections.
Some of these backgrounds describe the noncompact part of a
ten-dimensional critical superstring, while some others describe
lower-dimensional non-critical superstrings. The way we will identify
the correct superstring is by looking at the Ricci scalar of the
backgrounds, which vanishes for the backgrounds being a part of a
compactification.

The coefficients of the one-loop beta-function equations for the
conformal invariance of a sigma-model on a supercoset $G/H$ with
$\mathbb{Z}_4$ automorphism are proportional to the super Ricci tensor
of the supergroup $G$. This has been shown for the matter part of the
hybrid formalism in \cite{Berkovits:1999im,Berkovits:1999zq} (which is
identical to the matter part of the pure spinor action) and we will
show below that the same holds for the ghost part of the pure spinor
action. Whenever the supergroup $G$ is super Ricci flat (its dual
Coxeter number vanishes), the sigma-model is automatically conformally
invariant at one-loop. The supercosets describing $AdS_p\times S^p$
backgrounds are all super Ricci flat and therefore conformally
invariant. Moreover, since the $AdS_p$ and the $S^p$ part have the
same radii, their scalar curvatures have equal modulus but opposite
sign and hence the total scalar curvature of the background
vanishes. Since in these backgrounds the dilaton is constant and the
scalar curvature vanishes, the Weyl anomaly also vanishes and they
necessarily describe a part of a critical ten-dimensional
background. The compactified part has to be Ricci flat and preserve
minimal supersymmetry, hence a CY manifold of complex dimension $5-p$
would do the job and we can identify the full ten-dimensional
background as $AdS_p\times S^p\times CY_{5-p}$.

When the supergroup $G$ is not super Ricci flat (its dual Coxeter
number is nonvanishing), the GS sigma-model can be still shown to be
conformal at one-loop, as first discussed by Polyakov
\cite{Polyakov:2004br}. The intuitive reason, which we will explain
below, is that classical $\kappa$-symmetry of the GS action, which is
responsible for spacetime supersymmetry, is enough to ensure one-loop
conformal invariance. The scalar curvature of these backgrounds is
nonvanishing and we will argue that they describe a non-critical
superstring, along the lines of \cite{Polyakov:2004br}. However, we
will not compute the Weyl beta-function, in other words we are not
computing the central charge.  As a heuristic check of consistency, we
will just show that in all cases the naive central charge of the
matter plus ghost action that we get in the free field theory limit
vanishes. However, a precise computation of the central charge for the
non-critical case would require the ability to analyze strongly
coupled sigma-models.

Before going into the details, let us summarize the results. We
will collect first the ten-dimensional backgrounds and then the
non-critical ones. In all cases we have given both the classical
Green-Schwarz and the quantum pure spinor sigma
models\footnote{While the GS sigma-model always describes the full
superstring, some subtleties concern the pure spinor action. In
this latter case, it has been argued in \cite{Adam:2006bt} that
the non-critical pure spinor formalism describes the full
non-critical superstring spectrum. On the other hand, it might be
that the $AdS_p\times S^p\times CY_{5-p}$ lower-dimensional pure
spinor action is to be interpreted as the ``topological sector"
of the full ten-dimensional superstring compactified on CY. The
reason for this is that the cohomology of lower-dimensional pure
spinor theories in flat Minkowski describes the off-shell
multiplets of lower-dimensional supersymmetry
\cite{Berkovits:2005bt,Grassi:2005jz}, and the same structure
might carry on to other curved backgrounds.}.

\subsubsection*{Critical superstrings}

The following backgrounds are interpreted as the non-compact part
of a ten-dimensional type II background $AdS_p\times S^p\times
CY_{5-p}$.

\begin{itemize}
\item $AdS_2\times S^2$ with RR two-form flux, realized as
 \be
 AdS_2\times S^2: {PSU(1,1|2)\over U(1)\times
 U(1)},\label{ads2coset}
 \ee
is super Ricci flat. Therefore, it is the non-compact part of the
ten-dimensional type IIA background obtained by tensoring it with
a compact CY threefold as in \cite{Berkovits:1999zq}.
 \item $AdS_3\times S^3$ with RR three-form flux, realized as
 \be
 AdS_3\times S^3: {PSU(1,1|2)^2\over SO(1,2)\times
 SO(3)},\label{ads3coset}
 \ee
is super Ricci flat as well. Therefore, it is the noncompact part
of the ten-dimensional type IIB background obtained by tensoring
it with a compact CY twofold.
 \end{itemize}

\subsubsection*{ Non-critical superstrings}

The following backgrounds are interpreted as non-critical
superstrings.
The $AdS_{2n}$ backgrounds with $2n$ units of RR-flux, which we
realized as
\bea
 AdS_2:&{Osp(2|2)\over SO(1,1)\times SO(2)}\nn\\
 AdS_4:&{Osp(2|4)\over SO(1,3)\times SO(2)}\\
 AdS_6:&{F(4)\over SO(1,5)\times SL(2)}\nn
\eea
describe type IIA non-critical superstrings in $2n$
dimensions.\footnote{In addition, the $AdS_2$
non-critical background can be realized as an $Osp(1|2)/SO(2)$ supercoset. The
classical GS sigma-model for this supercoset is well defined
\cite{Verlinde:2004gt}. However, as we will see, in its quantum
realization as a pure spinor superstring all the correlation
functions vanish. We do not know how to interpret this fact.}

\subsection{One-loop beta-function}
\label{validity}

Before verifying that the various backgrounds are conformally
invariant at one-loop, we would like to warn the reader about the
validity of such computations.

The backgrounds of the ten-dimensional critical superstring can be
usually considered in the regime in which the spacetime curvature is
very small. In this regime supergravity is a good approximation. In
the example of $AdS_5\times S^5$, this is the limit where the radius
of $AdS$ is very large. Since the radius corresponds to the inverse
coupling of the sigma-model, the small curvature limit is realized as
the weak coupling regime of the sigma-model. Thus, in this case it
makes sense to study the conformal invariance of the worldsheet theory
order by order in the sigma-model perturbation theory and one finds that
one-loop conformal invariance requires an on-shell supergravity
background (small curvature limit). Higher loops in the sigma-model
describe higher curvature corrections to the supergravity equations of
motion.

In the case of non-critical superstrings things are typically
different. Namely, the curvature is always at string scale. In
fact, as we already mentioned, there is no regime in which
non-critical supergravity (one-loop perturbation theory in the
sigma-model) provides a reliable description of the
spacetime.\footnote{By non-critical supergravity, as will be
clarified below, we mean lower-dimensional supergravity with a
cosmological constant term, fixed at string scale value.}
Therefore the sigma-models that we described in the previous
sections are typically strongly coupled two-dimensional field
theories. In particular, they are understood to be living at a
fixed point of the worldsheet RG flow.

With this caveat in mind, in this section we will check that these
sigma-models are conformally invariant at one-loop. We take this
as an evidence for the existence of these theories, while we leave
for a future analysis a proof of conformal invariance at all
orders in perturbation theory.

We review the computation of the one-loop conformal beta-function in
the GS sigma-models \cite{Polyakov:2004br}. We will not consider
the one-loop beta-function for the Weyl anomaly. We will see then
that the contribution of the bosonic part to the one-loop
effective action precisely cancels the contribution of the
fermionic part, proving one-loop conformal invariance. This is due
to the fact that $\kappa$-symmetry fixes the number of physical
bosons equal to the number of physical fermions, implementing
therefore spacetime supersymmetry. A sigma-model on a
$d$-dimensional background has $d-2$ physical bosonic degrees of
freedom in both left and right moving sectors. $\kappa$-symmetry
requires that the number of physical fermions should also be $d-2$
in both the left and right moving sectors, which fixes the total
number of real spacetime supersymmetries to $4(d-2)$. This gives
us sixteen supersymmetries in six dimensions and eight
supersymmetries in four dimensions (which is the same number as
required in type II compactification on CY). In two dimensions,
however, since $\kappa$-symmetry removes all the fermionic degrees
of freedom, we can have more possibilities, namely two or four. We
will argue in the next section what happens in this last
case.

Here we review the computation of the one-loop beta-function of
the $AdS_4$ coset $\frac{OSp(2|4)}{SO(3, 1) \times SO(2)}$
performed in \cite{Polyakov:2004br} adapting it to the notations
used in this paper. We begin with the action
(\ref{eq:generic-GS-coset-action}) which in the conformal gauge
reads
\begin{equation}
  S = \frac{1}{\lambda^2} \int d^2 \sigma \left\langle \frac{1}{2}
  J_2 \bar J_2 + \frac{1}{4} J_1 \bar J_3 - \frac{1}{4} J_3 \bar J_1
  \right\rangle \ ,
\end{equation}
in which the coupling $\lambda$ is written explicitly. The
dependence of the sigma-model coupling $\l$ on the string coupling
$g_S$ and the RR flux $N_c$ is given by
 \be {1\over\l^2}=g_SN_c \ .
 \ee
We consider the quantum fluctuations $X \in osp(2|4)$ around the
classical background $\tilde g$ such that $g = \tilde g e^{\lambda
  X}$. The currents are given by
\begin{equation}
  J = e^{-\lambda X} \partial e^{\lambda X} + e^{-\lambda X} \tilde J
  e^{\lambda X} \ ,
\end{equation}
and the corresponding equation for the right-moving current, where
$\tilde J = \tilde g^{-1} \partial \tilde g$ and similarly for the
right-movers. As argued in \cite{Berkovits:1999zq} the gauge $X
\in \mathcal{G} \backslash \mathcal{H}_0$ can be chosen. The
one-loop beta-function is obtained from the second order expansion
in $\lambda$ of the action. By computing the first order
expansion, integrating by parts and making use of the
Maurer-Cartan equations to express the derivatives of the currents
$J_1$ and $J_3$ in terms of the commutators of currents one gets
\begin{eqnarray}
  S_1 & = & \frac{1}{\lambda} \int d^2 z \bigg\langle \frac{1}{2} \partial
  X_2 \bar J_2 + \frac{1}{2}J_2 \bar \partial X_2 - \frac{1}{2} ([J_0,
  \bar J_2] - [J_2, \bar J_0] - [J_1, \bar J_1] + [J_3, \bar J_3]) X_2
  + \nonumber\\
  && {} +[J_2, \bar J_1] X_1 - [J_3, \bar J_2] X_3 \bigg\rangle \
  ,
\end{eqnarray}
where we dropped the tilde on the background currents for simplicity of
notation. The second variation is then computed and as in
\cite{Polyakov:2004br} it is convenient to restrict the
computations to backgrounds with $J_1 = J_3 = \bar J_1 = \bar J_3
=0$ since $\kappa$-symmetry guarantees that the beta-function
associated with the other terms in the action will be equal to the
beta-function of the term $J_2 \bar J_2$. In such backgrounds the
action for the $X$ fields reduces to
\begin{eqnarray}
  S_X & = & \int d^2 z \Big\langle \partial X_2 \bar \partial X_2 +
  [\bar J_0, X_2] \partial X_2 + [J_0, X_2] \bar \partial X_2 - [\bar
  J_2, X_2] [J_2, X_2] + [\bar J_0, X_2] [J_0, X_2] - \nonumber\\
  && {} - [J_2, X_1] \bar \partial X_1 - [\bar J_2, X_3] \partial X_3
  - [J_2, X_1] [\bar J_0, X_1] - 2 [J_2, X_1] [\bar J_2, X_3] -
  \nonumber\\
  && {} - [\bar J_2, X_3] [J_0, X_3] + O(\lambda) \Big\rangle \ .
\end{eqnarray}
In order to compute the one-loop quantum corrections to the $J_2 \bar
J_2$ term we write the relevant parts of the action in terms of the
structure constants and the Cartan metric
\begin{eqnarray}
  S_X & = & \int d^2 z ( \eta_{a b} \partial X_2^a \bar \partial
  X_2^b + \eta_{[ef] [gh]} f^{[ef]}_{a b} f^{[gh]}_{c d} J_2^a
  \bar J_2^c X_2^b X_2^d + \eta_{\hat \alpha \alpha} f^{\hat
  \alpha}_{a \beta} J_2^a X_1^\beta \bar \partial X_1^\alpha +
  \nonumber\\
  && {} + \eta_{\alpha \hat \alpha} f^\alpha_{a \hat \beta} \bar
  J_2^a X_3^{\hat \beta} \partial X_3^{\hat \alpha} + 2 \eta_{\hat
  \alpha \alpha} f^{\hat \alpha}_{a \beta} f^\alpha_{b \hat \beta}
  J_2^a \bar J_2^b X_1^\beta X_3^{\hat \beta} + \dots ) \ .
\end{eqnarray}

Upon substituting the structure constants of $OSp(2|4)$ (see
Appendix C) we get
\begin{eqnarray}
  S_X & = & \int d^2 z [\eta_{a b} \partial X_2^a \bar \partial X_2^b
  - (\delta_{a c} \delta_{b d} - \delta_{a d} \delta_{b c}) J_2^a \bar
  J_2^c X_2^b X_2^d - J_2^a \bar X_{1 \beta} {(\gamma_a)^\beta}_\alpha
  \bar \partial X_1^\alpha - \nonumber\\
  && {} - \bar J_2^a \bar X_{3 \hat \beta} {(\gamma_a)^{\hat
  \beta}}_{\hat \alpha} \partial X_3^{\hat \alpha} - J_2^a \bar J_2^b
  \bar X_{1 \beta} {(\gamma_a \gamma_b)^\beta}_{\hat \beta} X_3^{\hat
  \beta} + \dots ] \ ,
\end{eqnarray}
where $X_1$ and $X_3$ satisfy the Majorana conditions $\bar X_{1
\beta} = X_1^\alpha C_{\alpha \beta}$ and $\bar X_{3 \hat \beta} =
X_3^{\hat \alpha} C_{\hat \alpha \hat \beta}$.  The
$\kappa$-symmetry is most conveniently fixed as in
\cite{Polyakov:2004br}
\begin{displaymath}
  J_2^a \gamma_a = \sqrt{J_2^a \bar J_2^a} \gamma_+ \ , \quad
  \bar J_2^a \gamma_a = \sqrt{J_2^a \bar J_2^a} \gamma_- \ , \quad
  X_{1, 3} = (J_2^a \bar J_2^a)^{-1/4} Y_{1, 3} \ ,
\end{displaymath}
where after fixing the residual conformal symmetry the bosonic
indices run only on the two transverse directions and the Majorana
spinors $Y_{1, 3}$ satisfy the light-cone constraints $\gamma_+
Y_3 = \gamma_- Y_1 = 0$. The free field OPEs for the fluctuations
now take the form
\begin{eqnarray}
  X_2^a (z, \bar z) X_2^b (0, 0) & \sim & -\frac{1}{4 \pi} \eta^{a b} \log
  |z|^2 \ ,\\
  \bar Y_{1\alpha} (z, \bar z) Y_1^\beta (0, 0) & \sim & -\frac{1}{8
  \pi z} {(\gamma_-)^\beta}_\alpha \ ,\\
  \bar Y_{3 \hat \alpha} (z, \bar z) Y_3^{\hat \beta} (0, 0) & \sim &
  -\frac{1}{8 \pi \bar z} {(\gamma_+)^{\hat \beta}}_{\hat \alpha} \ .
\end{eqnarray}

\FIGURE[tb]{
\includegraphics{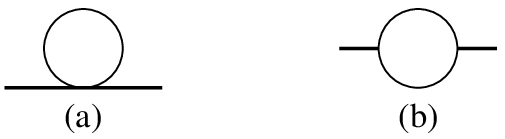}
\caption{The bosonic (a) and the fermionic (b) diagrams contributing
to the one-loop beta-function} \label{fig:diagrams}
}

Thus the bosonic one-loop correction to the effective action
coming from figure~\ref{fig:diagrams}(a) is
\begin{displaymath}
  -\frac{1}{2 \pi} J_2^a \bar J_2^a \log \frac{\Lambda}{\mu} \ ,
\end{displaymath}
where the UV cut-off is $|z| = 1 / \Lambda$ and the IR one is $|z|
= 1 / \mu$. Similarly, the one-loop fermionic correction to the
$\left\langle J_2 \bar J_2 \right\rangle$ comes from the diagram
in figure~\ref{fig:diagrams}(b) and evaluates to
\begin{displaymath}
  \frac{1}{2 \pi} J_2^a \bar J_2^a \log \frac{\Lambda}{\mu}
\end{displaymath}
so the total one-loop correction to $\left\langle J_2 \bar J_2
\right\rangle$ vanishes. The gauge symmetry of the sigma-model
guarantees that no terms involving $J_0$ appear and unless
$\kappa$-symmetry does not hold quantum mechanically, the
$\left\langle J_1 \bar J_3 \right\rangle$ and $\left\langle J_3
\bar J_1 \right\rangle$ are not corrected as well to one-loop.

A note on the difference between the $AdS_p \times S^p$ background
and the $AdS_p$ backgrounds is in order. For the former it was
found that super-Ricci flatness of the group $G$ in the coset
$G/H$ was a sufficient condition for one-loop conformal invariance
of $AdS_p \times S^p$ backgrounds \cite{Berkovits:1999zq}. However,
as we see here, it is not a necessary condition as demonstrated by
the latter case since the Maurer-Cartan equations allow to relate
contributions to the $J_1 \bar J_3$ and $J_3 \bar J_1$
beta-functions to the $J_2 \bar J_2$ one leading to the vanishing
of the beta-function.

In the $AdS_2$ case perturbative conformal invariance is trivial
because as discussed in subsection \ref{sec:non-critical-AdS_2}
there are no propagating degrees of freedom after fixing the
$\kappa$-symmetry.

{\it Pure spinor beta-functions}

Let us comment on the computation of the beta-function in the pure
spinor formalism in the background field method. Unlike the
light-cone GS formalism, we work covariantly at all stages. The
matter part of the action is identical to the corresponding
formulation of the hybrid superstring on a supercoset with $\ZZ_4$
automorphism, which was considered in \cite{Berkovits:1999zq}.
However, when the supergroups $G$ have nonzero dual Coxeter
number, as in the non-critical backgrounds, the various terms
rearrange differently.

The contribution to the one-loop effective action coming from the
pure spinor sector was considered in \cite{Vallilo:2002mh} for
$AdS_5\times S^5$. It consists of two terms. The first term is
obtained by expanding the ghost action $\frac{1}{\lambda^2}\int
d^2z\,\langle N\bar J_0+\bar N J_0\rangle$ to the second order in the
fluctuations of the gauge current $J_0$. The trilinear couplings
 \bea
\int d^2z\, \langle \tilde N\left([\bar \partial X_2,X_2]+[\bar
\partial X_1,X_3]+[\bar
 \partial X_3,X_1]\right)\\
 +\tilde{\bar N}\left([\partial X_2,X_2]+[\partial X_1,X_3]+[\partial
 X_3,X_1]\right)\rangle,
 \eea
 generate the term $\langle \tilde N\tilde {\bar N}\rangle$ in
 the action through the fish diagram in Fig.\ref{fig:diagrams}(b)
 \bea
 {1\over 8\pi}{\rm log}{\Lambda\over \mu} \tilde N^{[ij]}\tilde{\bar N}^{[kl]}\left(
 4R_{[ij][kl]}(G)-4R_{[ij][kl]}(H)\right).\label{ghostfish}
 \eea

As explained in \cite{Vallilo:2002mh}, there is a second
contribution to the one-loop effective action in the ghost sector,
coming from the operator ${\cal O}(z,\bar z)=\langle N\bar
N\rangle$, which couples the pure spinor Lorentz currents to the
spacetime Riemann tensor. The marginal part of the OPE of ${\cal
O}$ with itself generates at one-loop the following contribution
to the effective action
 \be
 {1\over4\pi}\int d^2z\int d^2w \langle {\cal O}(z,\bar z){\cal O}(w,\bar
 w)\rangle={1\over2\pi}{\rm log}{\Lambda\over \mu}R_{[ij][kl]}(H)\int
 d^2z \tilde N^{[ij]}\tilde{\bar N}^{[kl]},
 \ee
which cancels the term proportional to $R_{[ij][kl]}(H)$ in
(\ref{ghostfish}). So we are left with the following ghost
contribution to the one-loop effective action in the ghost sector
\bea
 {1\over2\pi}{\rm log}{\Lambda\over \mu} \tilde N^{[ij]}\tilde{\bar N}^{[kl]}
 R_{[ij][kl]}(G) \ ,
 \eea
where the explicit expression of the super Ricci tensor of the
supergroup in terms of the structure constants is explained in the
appendix. In the $AdS_p\times S^p$ cases \cite{Berkovits:1999zq},
in which the supergroup $G$ is super Ricci flat, each coupling in
the effective action vanishes by itself, all of them being
separately proportional to the dual Coxeter number of the
supergroup $G$. However, in the non-critical superstrings, in
which the dual Coxeter number of $G$ is nonzero, even if the
single terms do not vanish separately, one expects that by making
use of Ward identities they give a total vanishing contribution.
We just mention that the nontrivial cancellation between the
various couplings in the effective action is precisely what
happens in the GS computation above. In that case, in the physical
gauge there are no ghosts. The bosonic and fermionic part of the
beta-function are both non-vanishing (if the dual Coxeter number
of $G$ is non-vanishing), however they exactly cancel due to
$\kappa$-symmetry and using the Maurer-Cartan equations. In the
pure spinor formulation, the BRST symmetry plays the role of the
$\kappa$-symmetry and we have the ghost contribution as well
because we work covariantly. At the end of the day, the physical
reason for the vanishing of the beta-function would be again
spacetime supersymmetry. We leave the proof of one-loop conformal
invariance of the pure spinor action in the case of nonzero dual
Coxeter number for a future analysis.

\section{The various backgrounds}
\label{examples}

In this section we give some details on the various backgrounds
for which our general construction can be applied. They are all
realized as supercosets $G/H$, where the gauge symmetry $H$ is the
invariant locus of a $\ZZ_4$ automorphism of $G$. The details of
the supergroups and their structure constants are collected in
Appendix \ref{superapp}.

The existing literature on type II superstrings on $AdS$
backgrounds with RR flux, realized as sigma-models on supercosets,
is vast. The Green-Schwarz superstring on $AdS_p\times S^p$ has
been first constructed in the case $p=5$ \cite{Metsaev:1998it} and
subsequently in the compactified cases $p=3$
\cite{Pesando:1998wm,Rahmfeld:1998zn} and $p=2$
\cite{Zhou:1999sm}. The Green-Schwarz non-critical superstring on
$AdS_2$ has been proposed in \cite{Verlinde:2004gt}\footnote{As
explained below, the superstring based on the $Osp(1\vert2)$
supercoset in \cite{Verlinde:2004gt} is different from ours, based
on the $Osp(2\vert2)$ supercoset.}, while the non-critical $AdS_4$
has been discussed in \cite{Polyakov:2004br}. The type II pure
spinor action for $AdS_5\times S^5$ has been introduced in
\cite{Berkovits:2000fe,Berkovits:2002zk}. The hybrid formalism for
the critical cases $p=2,3$, whose matter part is similar to the
matter part of our pure spinor sigma-models, have been discussed
in \cite{Berkovits:1999im,Berkovits:1999zq}. The type II pure
spinor action for non-critical $AdS_4$ has been proposed in
\cite{Adam:2006bt}. The analysis of conformal invariance of such
superstring sigma-models in the Green-Schwarz
\cite{Polyakov:2004br}, hybrid
\cite{Berkovits:1999im,Berkovits:1999zq} and pure spinor
\cite{Vallilo:2002mh,Berkovits:2004xu} formulations has received
some attention as well.

The proof of the classical integrability of the Green-Schwarz
sigma-model on $AdS_5\times S^5$ \cite{Bena:2003wd} has boosted
the attention on the integrability of sigma-models on supercosets
\cite{Young:2005jv,Kagan:2005wt,Babichenko:2006uc,Kluson:2006ij}.
Classical integrability of the type II Green-Schwarz superstring
on various critical and non-critical backgrounds has been further
studied in \cite{Polyakov:2004br,Chen:2005uj,Polyakov:2005ss}. The
integrability of the pure spinor superstring on $AdS_5\times S^5$
has been proven first from the classical point of view
\cite{Vallilo:2003nx,Berkovits:2004jw} and afterwards quantum
mechanically \cite{Berkovits:2004xu}.

\subsection{Non-critical $AdS_2$} \label{sec:non-critical-AdS_2}

The type IIA non-critical superstring on $AdS_2$ with RR two-form
flux is realized as the supercoset $Osp(2|2)/SO(1,1)\times SO(2)$.
The $Osp(2|2)$ supergroup has four bosonic generators $(\bf
E^\pm,\bf H,\bf \tilde H)$ and four fermionic ones $(\bf Q_\a,\bf
Q_{\ah})$. The index $a=\pm$ denotes the spacetime light-cone
directions. The supercharges are real two-dimensional MW spinors,
the index $\a=1,2$ counts the ones with left spacetime chirality
and the index $\ah=\hat 1,\hat 2$ counts the ones with right
spacetime chirality (note that in the two-dimensional superstring
$\a,\ah$ are not spinor indices but just count the multiplicity of
spinors with the same chirality). To obtain $AdS_2$, we quotient
by $\mathbf{H}$ and $\tilde \HH$, which generate respectively the
$SO(1,1)$ and $SO(2)$ transformations. The $Osp(2|2)$ superalgebra
and structure constants are listed in an appendix. The left
invariant form $J=G^{-1}dG$ is expanded according to the grading
as
 \bea
 J_0=J^H\HH+J^{\tilde
 H}\tilde\HH,\quad
 J_1=J^\a\QQ_\a,\quad J_2=J^a\EE_a,\quad J_3=J^{\ah}\QQ_{\ah}.
 \eea
and the definition of the supertrace is
 \bea
 \langle
 \EE_a\EE_b\rangle=\delta_{a}^{+}\delta_{b}^{-}+\delta_{a}^{-}\delta_{b}^{+},
 \qquad \langle
 \QQ_\a\QQ_{\ah}\rangle=\delta_{\a\ah},\label{super2}
 \eea
whose details are given in an appendix.

{\it $\kappa$-symmetry of the GS action in two dimensions}

The Green-Schwarz sigma-model on $AdS_2$ is given by
(\ref{eq:generic-GS-coset-action}). Let us discuss its
$\kappa$-symmetry. If we want to define a classical Green-Schwarz
superstring in a flat background, it is well known that only in
$d=3,4,6,10$ does $\kappa$-symmetry exist. This is due to the
existence of the Fierz identities in the gamma matrix algebra, which
are needed to define the WZ term. Even if the GS superstring does not
exist in flat two dimensions, it does exist on a two-dimensional $AdS$
background with RR two-form flux. The RR flux makes it possible to
construct a WZ term, as we showed above. An example of these
two-dimensional $AdS$ sigma-models has been discussed in
\cite{Verlinde:2004gt}. There is, however, a substantial difference
between the usual higher-dimensional $\kappa$-symmetry and the
two-dimensional one. In higher dimensions this gauge symmetry is
reducible. As a result, it removes only half of the $\theta$'s from
the classical spectrum. In two dimensions, instead, it is not
reducible and it removes all the $\theta$'s. This fact is expected
from two-dimensional on-shell supersymmetry. In the light-cone gauge
the two bosonic coordinates are removed from the spectrum, leaving no
bosonic degrees of freedom. It is a necessary requirement then that
the worldsheet fermionic symmetry remove all the fermionic degrees of
freedom as well, and not just half as in higher dimensions. Let us see
how this works in detail. We will first briefly review the
reducibility of the ten-dimensional $\kappa$-symmetry on $AdS_5 \times
S^5$ and then show that in $AdS_2$ it is not reducible.

In $AdS_5\times S^5$ the fermionic coordinates $\theta^\a,
\theta^\ah$ transform under $\kappa$-symmetry as follows
\cite{Metsaev:1998it}. Working out the algebra in the
transformations (\ref{kappasym}), we find
 \bea
 \delta \theta^\a=(A_z)^\a_\bh
 (\kappa^z)^\bh,\label{thetasym}\qquad
 \delta \theta^\ah=(A_{\bar z})^\ah_\b (\kappa^{\bar z})^\b,
 \eea
where we picked the worldsheet conformal gauge. Here, $\kappa^\bh$
and $\kappa^\b$ are the fermionic gauge parameters; each one of
them has sixteen real components of a MW spinor and a holomorphic
or anti-holomorphic vector index. The crucial point is the
presence of the (field dependent) matrices $A_z,A_{\bar z}$, whose
explicit form is $(A_z)^\a_\bh= J^m_z(\gamma_m)_{\ah\bh}P^{\a\bh}$
and $(A_{\bar z})^\ah_\b=- J^m_{\bar
z}(\gamma_m)_{\a\b}P^{\b\ah}$, where $P^{\a\ah}=\half
\delta^{\a\ah}$ is the RR five-form flux. Due to the Virasoro
constraints $J^m_zJ_{zm}=0=J^m_{\bar z}J_{\bar zm}$, these two
matrices are not invertible, and in fact they have rank eight,
rather than sixteen (it is easy to see this, e.g. because they are
nilpotent). As a result, only eight degrees of freedom can be
removed by this gauge symmetry. The following choice of the gauge
parameters
 \bea
 \kappa_{(3)}^z=-[J_{2z},\e^z_{(1)}],\label{specialk}\qquad
 \kappa_{(1)}^{\bar z}=[J_{2\bar z},\e^{\bar z}_{(3)}],
 \eea
in fact, gives $\delta \theta^\a=\delta \theta^\ah=0$. Similar
considerations apply to the cases $d=4,6$.

In two dimensions, the spacetime fermionic coordinates are MW
spinors with one real component. Therefore, the (non-invertible)
matrices $(A_z)^\a_\bh$ and $(A_{\bar z})^\ah_\b$ of
(\ref{thetasym}) are replaced by ordinary functions. If we
consider the case of the supercoset $Osp(2|2)/SO(2)^2$, the
transformations (\ref{kappasym}) read\footnote{The indices
$(\a,\ah)$ are not spinor indices, but simply label the
multiplicity of the fermionic coordinates with the same chirality.
The case $Osp(1|2)/SO(2)$ considered in \cite{Verlinde:2004gt} is
recovered by simply dropping the indices $\alpha$, $\hat \alpha$.}
 \bea
 \delta \theta^\a=-E^+_z\kappa^{z\a},\qquad
 \delta \bar \theta^\ah=-E^-_{\bar z}\bar\kappa^{\bar
 z\ah}.\label{kappasym2}
 \eea
The special choice (\ref{specialk}) still gives
$\delta\theta^\a=\delta\bar\theta^\ah=0$. However, now each
coordinate can be gauged away independently and the gauge
parameter is not multiplied by a nilpotent matrix, but rather by
just an ordinary function. Hence, $\kappa$-symmetry in two
dimensions is not reducible and gauges away all the fermionic
coordinates. The counting of classical degrees of freedom thus
agrees with target space supersymmetry, leaving an empty classical
spectrum.

This can be understood at the level of the constraint algebra as
well. Consider first the ten-dimensional GS action $S_{GS}$. Since
the conjugate momenta $p$'s to the $\theta$'s do not involve time
derivative, we find the so called GS constraint
 \bea
 d_\a=p_\a-{\delta S_{GS}\over
 \delta\dot\theta^\a}\approx0,\label{ddef}\qquad
 \bar d_\ah=\bar p_\ah-{\delta S_{GS}\over \delta\dot{\bar
 \theta}^\ah}\approx0,\nn
 \eea
that satisfy the classical algebra
 \bea
 \{d_\a,d_\b\}=-\Pi^m(\gamma_m)_{\a\b},\qquad
 \{\bar d_\ah,\bar d_\bh\}=-\bar
 \Pi^m(\gamma_m)_{\ah\bh}.\label{dalgebra}
 \eea
Each GS constraint is a ten-dimensional MW spinor with sixteen
real components. Eight of these components are first class
constraints that generate the $\kappa$-symmetry transformations in
(\ref{thetasym}). The remaining eight components are second class
constraints. However, the first and second class constraints are
mixed and it is not possible to disentangle them in a manifestly
covariant way. This can be seen in the algebra (\ref{dalgebra}),
since the right hand sides are not vanishing but they are
nilpotent matrices on the Virasoro constraints $\Pi^m\Pi_m=0=\bar
\Pi^m\bar \Pi_m$. In other words, the right hand sides of the
constraint algebra are projectors.

In two dimensions the situation is simpler. We still have two sets
of constraints (\ref{ddef}), whose algebra now reads
 \bea
 \{d_\a,d_\b\}=-\delta_{\a\b}E^+_z,\qquad
 \{\bar d_\ah,\bar d_\bh\}=-\delta_{\ah\bh}\bar
 E^-_{\bar z}.\label{dalgebra2}
 \eea
The claim is that all the GS constraints are now first class. In
fact, in two dimensions the Virasoro constraints are
$E^+_zE^-_z=0=\bar E^+_{\bar z}\bar E^-_{\bar z}$, because we just
have the two light-cone directions. Therefore a consistent
solution to the Virasoro constraint is $E^+_z=0=\bar E^-_{\bar
z}$. As a result, the algebra of the fermionic constraints now
closes on first class constraints, namely it is weakly zero. So
there are no second class constraints. Now all the $d$'s are
generators of the $\kappa$-symmetry (\ref{kappasym2}), by which we
can remove all the fermionic variables.

{\it Pure spinor sigma-model}

The action of the pure spinor sigma-model is given by
(\ref{eq:AdS-ps-action}), where the pure spinor
$\beta\gamma$-system is defined according to (\ref{purepara}). The
left and right moving pure spinors $\l^\a$ and $\bar \l^\ah$
satisfy the pure spinor constraints (\ref{2dpurity})
 \bea
 \l^\a\delta_{\a\b}\l^\b=0,\qquad \bar \l^\ah\delta_{\ah\bh}\bar\l^\bh=0.
 \eea
This is the pure spinor space for two-dimensional type II
non-critical superstrings that was introduced in
\cite{Grassi:2005sb,Wyllard:2005fh,Adam:2006bt}, to which we refer for
more details (the different notations are explained in the footnote
\ref{ourfoot}). 

{\it Holography and isometries}

The type IIA superstring on $AdS_2$ has a natural candidate for
the gauge theory dual, living on the boundary of $AdS_2$. It is a
superconformal matrix quantum mechanics with global symmetry group
$Osp(2|2)$, corresponding to the global symmetries of the
worldsheet theory. The gauge theory is the worldvolume theory
living on a stack of many D-particles of the type IIA superstring
and is described by a Marinari-Parisi quantum mechanics.

A new feature of holography in our setup is that not all the
global symmetries of the dual gauge theory come from isometries of
the closed string background. In the usual example of $AdS/CFT$,
the duality is between type IIB superstring theory on $AdS_5\times
S^5$ background and ${\cal N}=4$ SYM theory in four dimensions.
The $SU(4)$ R-symmetry of the four-dimensional gauge theory
corresponds on the closed string side to the isometry group
$SO(6)\simeq SU(4)$ of the compact manifold $S^5$. Consider now
the $AdS_2$ non-critical superstring. It is invariant with respect
to a global $SO(2)$ symmetry, generated by $\widetilde
\HH$.\footnote{Even if we eventually quotient by this generator,
the global symmetry of the superstring is the full $Osp(2|2)$. In
the same way, in the $AdS_5\times S^5$ example, the superstring is
invariant with respect to the full supergroup $PSU(2,2|4)$.} This
symmetry corresponds again to the R-symmetry on the gauge theory
side. However, this rotation does not correspond to any isometry
of the closed string background, still it is a global symmetry of
the closed string theory. The interpretation of this kind of
non-geometric symmetry would fit with the intuition coming from
the holography in the case of gauged supergravity. In the gravity
spectrum, in that case, there are some additional gauge fields
which couple to dual gauge theory operators and explain the extra
gauge symmetry. On the string side there are some vertex operators
with the correct R-charge assignment, that we could scatter to
reproduce the gauge theory computation.

{\it The $Osp(1|2)$ supercoset}

We would like to make some comments here on a different realization of
type IIA non-critical superstrings on $AdS_2$ background that was
proposed by Verlinde \cite{Verlinde:2004gt}. The action is based on
the supercoset $Osp(1|2)/SO(1,1)$. The algebra of $Osp(1|2)$ can be
easily obtained from the one of $Osp(2|2)$ (that we list in the
appendix) by dropping the indices $\a,\ah$, thus removing half of the
fermionic generators and discarding the bosonic generator
$\mathbf{\tilde H}$. The sigma-model constructed in
\cite{Verlinde:2004gt} can be recast in the usual Green-Schwarz like
form (\ref{eq:generic-GS-coset-action}) using the grading zero
Maurer-Cartan identity relating the exterior product of the bosonic
Cartan one-forms with that of the fermionic one-forms. Then, one can
apply the machinery developed in this paper to prove that the
Green-Schwarz sigma-model still has an infinite number of nonlocal
conserved charges, precisely of the form given in
(\ref{flatsolution}). As a result, the classical Green-Schwarz
superstring on the supercoset $Osp(1|2)$ is well defined and
integrable.

The pure spinor sigma-model is still given formally by the action
(\ref{eq:AdS-ps-action}). But when we try to identify the pure
spinor variables, we encounter the following feature. The left- and
right-moving pure spinors $\l$ and $\bar \l$ are Weyl spinors of
opposite chirality which satisfy the pure spinor
constraint
 \bea
    \l^2=0, \qquad \bar \l^2=0.\label{puri21}
 \eea
This fits in the general discussion of section \ref{puresection},
by noting that the supersymmetry algebra is now generated by the
superderivatives $d$ and $\bar d$ satisfying
 \bea
 \{d,d\}=-E^+_z,\qquad \{\bar d,\bar d\}=-\bar E^-_{\bar z}.
 \eea
The solution of the $Osp(1|2)$ pure spinor constraint (\ref{puri21})
requires that $\l=\psi_1\psi_2$ using the two Grassmann odd fields
$\psi_1,\psi_2$.

In the pure spinor sigma-model, the pure spinor variables are
interpreted as the ghosts. Consider for simplicity the left sector
only (the closed string is the product of the left and right
sectors). Then, the physical cohomology is described by operators
of ghost number one and weight zero, which in this case are ${\cal
U}=\l A(\theta,x^\pm)$, for a generic superfield $A$ depending on
the zero modes only. On general grounds, the prescription for the
tree level amplitudes in the pure spinor formalism requires the
insertion of three unintegrated vertex operators of ghost number
one
 \bea
 {\cal A}&=&\langle {\cal U}^{(1)} {\cal U}^{(2)}{\cal
 U}^{(3)}\ldots\rangle_{CFT},
 \eea
where the dots stand for a generic product of integrated vertex
operators.\footnote{The generalization of the ten-dimensional
saturation rule to the non-critical superstring was briefly
discussed in \cite{Adam:2006bt}.} However, all of these tree-level
amplitudes include products of three pure spinors which vanish
due to the pure spinor constraint.

\subsection{Non-critical $AdS_4$}
\label{ads4non}

The non-critical type IIA superstring on $AdS_4$ with RR four-form
flux is realized as a sigma-model on the $Osp(2|4)/SO(1,3)\times
SO(2)$ supercoset. The $Osp(2|4)$ superalgebra and structure constants
are discussed in the appendix. The bosonic generators are the
translations $\mathbf{P}_a$, the $SO(1,3)$ generators
$\mathbf{J}_{ab}$, for $a,b=1,\ldots,4$ and the $SO(2)$ generator
$\mathbf{H}$.  The fermionic generators are the supercharges
$\mathbf{Q}_\a,\mathbf{Q}_\ah$, where $\a,\ah=1,\ldots,4$ are
four-dimensional Majorana spinor indices. We have thus ${\cal N}=2$
supersymmetry in four dimensions. The charge assignment of the
generators with respect to the $\ZZ_4$ automorphism of $Osp(2|4)$ can
be read from the Maurer-Cartan one forms
  \bea
 J_0=J^{ab}{\mathbf{J}_{ab}}+J^{H}\HH,\quad
 J_1=J^\a\QQ_\a,\quad J_2=J^a{\mathbf{P}_a},\quad J_3=J^{\ah}\QQ_{\ah}.
 \eea

The non-critical Green-Schwarz sigma-model on $AdS_4$ was first
introduced in \cite{Polyakov:2004br,Adam:2006bt}. Again, it is
given by (\ref{eq:generic-GS-coset-action}), with the appropriate
definitions of the supertrace
 \bea
 \langle
 {\mathbf{P}_a}{\mathbf{P}_b}\rangle=\eta_{ab},
 \qquad \langle
 \QQ_\a\QQ_{\ah}\rangle=2\tilde C_{\a\ah},\label{super4}
 \eea
where $\tilde C_{\a\ah}$ is an antisymmetric matrix numerically
given by the four-dimensional charge conjugation matrix.

The pure spinor sigma-model, which was first introduced in
\cite{Adam:2006bt}, is given by (\ref{eq:AdS-ps-action}), where
the pure spinor $\beta\gamma$-system is defined according to
(\ref{purepara}). The left and right moving pure spinors $\l^\a$
and $\bar \l^\ah$ are four-dimensional Dirac spinors, satisfying
the pure spinor constraints (\ref{4dpurity}). This is the pure
spinor space for four-dimensional type II non-critical
superstrings that was discussed in \cite{Adam:2006bt}, to which we
refer for further details. In the case in which the RR flux is
space-filling, there is a subtlety in the definition of the action,
in particular in the coupling $\langle d\bar d\rangle$. In ten
dimensions \cite{Berkovits:2002zk}, this part of the action
couples the RR superfield $P^{\b\bh}$ to the fermionic variables
$d_\b$ and $\bar d_{\bh}$ as simply $d_\b P^{\b\bh}\bar d_{\bh}$.
In the $AdS_4$ case, the relation between the RR superfield and
the four-form field strength is
 \bea
 P^{\a\bh}&=& {1\over 4!}g_S(\tilde C\gamma_{m_1\ldots
 m_4})^{\b\bh}F_{m_1\ldots m_4}=g_S N_c (\tilde
 C\gamma^5)^{\b\bh}.
 \eea
Since the RR bispinor is proportional to $\gamma^5$, it is a
pseudoscalar quantity. On the other hand, we want the worldsheet
action to be a spacetime scalar, therefore in the GS action
(\ref{eq:generic-GS-coset-action}) the correct coupling is
$d_{\a}(\gamma^5 P)^{\a\ah}\bar d_{\ah}$. Since $(\gamma_5)^2=-\Id$,
we can again relate the sigma-model on the supergroup with the
background fields as explained in the appendix. Notice that the
$\gamma_{D+1}$ is present in the $\langle d\bar d\rangle$ part of the
action whenever the RR flux is space-filling and the spacetime
dimension is even, because in this case the RR bispinor superfield is
proportional to the product of all the gamma matrices. In the
two-dimensional case, however, we did not underline this subtlety,
because we used a one-dimensional spinor notation.

The theory dual to this closed superstring is a strongly coupled
three-dimensional SCFT with ${\cal N}=2$ supersymmetry and $U(1)$
R-symmetry. Note that the R-symmetry is realized in a non-geometric
way on the string side. It would be interesting to identify the dual
to this string theory and to study how the holographic map works in
the non-critical string.

\subsection{Non-critical $AdS_5\times S^1$ with open strings}

The $AdS_5\times S^1$ background with five-form flux can be
realized as the supercoset
 \bea
 AdS_5\times S^1&=&{SU(2,2|2)\over SO(1,4)\times SO(3)}. \label{supercoset5}
 \eea
The type IIB superstring theory on this background is not expected to
be consistent. Even if it is one-loop conformally invariant, the
beta-function for the Weyl invariance should be nonzero
\cite{Klebanov:2004ya}. In this section, we will first describe the
closed superstring sigma-model on the supercoset
(\ref{supercoset5}). In Appendix \ref{sixsugra} we will speculate
about a possible realization of this background as a strongly coupled
fixed point without the need of adding open strings.

The bosonic subgroup of $SU(2,2|2)$ is $SO(2,4)\times SO(3)\times
U(1)$. It has nineteen bosonic generators: the five translations
$\mathbf{P}_a$ along $AdS_5$ and the translation $\mathbf{R}$ along the
circle $S^1$, the ten angular momenta $\mathbf{J}_{ab}$ in $AdS_5$ and
the three bosonic generators $\mathbf{T}_a'$ of $SO(3)$. The sixteen
fermionic generators are given by the two supercharges
$\mathbf{Q}_{\a\a'},\mathbf{Q}_{\ah\ah'}$, where the unprimed indices
$\a,\ah=1,\ldots,4$ are five-dimensional spinor indices in the
Majorana representation and the primed indices $\a',\ah'=1,2$ are
$SO(3)$ spinor indices. The superalgebra and its structure
constants are listed in Appendix \ref{superapp}. The grading assignment of
the generators can be read off the following Maurer-Cartan forms
  \bea
 J_0=J^{ab}{\mathbf{J}_{ab}}+J^{a'}{\mathbf{T}_{a'}},&\qquad& J_2=J^a{\mathbf{P}_a}+J^R{\mathbf{R}},
 \\
 J_1=J^{\a\a'}\QQ_{\a\a'},&\qquad&
 J_3=J^{\ah\ah'}\QQ_{\ah\ah'}.\nn
 \eea

The $\kappa$-symmetric Green-Schwarz sigma-model is again
constructed as in (\ref{eq:generic-GS-coset-action}), with the
appropriate definitions of the supertrace
 \bea
 \langle
 {\mathbf{P}_a}{\mathbf{P}_b}\rangle=-\eta_{ab},&\quad &\langle
 {\mathbf{R}\mathbf{R}}\rangle=1,\\
\langle
 \QQ_{\a\a'}\QQ_{\ah\ah'}\rangle=\tilde C_{\a\ah}\tilde C_{\a'\ah'},
 \eea
where $\tilde C_{\a\ah}$ and $\tilde C_{\a'\ah'}$ are
antisymmetric matrices numerically given by the charge conjugation
matrices of $SO(1,4)$ and of $SO(3)$. This Green-Schwarz action,
although in a slightly different form, was discussed in
\cite{Polyakov:2004br,Chen:2005uj}. The classical sigma-model
action is $\kappa$-symmetric and its one-loop conformal beta-function vanishes.

However, by looking at the non-critical supergravity equations of
motion \cite{Klebanov:2004ya}, we know that we have to add an open
string sector, namely space-filling D-branes, in order to properly
cancel the Weyl anomaly, which from the spacetime point of view is
encoded in the dilaton equation of motion. We review in Appendix
\ref{sixsugra} the target space computation of
\cite{Klebanov:2004ya}. The computation of the Weyl anomaly, which
fixes the radius of the $AdS$, is tantamount to the evaluation of the
central charge of the quantum sigma-model at the strongly coupled
fixed point. As in the other examples, we would need strong coupling
techniques to address this question, which are lacking at the moment.

In order to study the quantum sigma-model, we can introduce the
pure spinor formulation of the supercoset by considering the
action (\ref{eq:AdS-ps-action}), where the pure spinor
$\beta\gamma$-system is defined according to (\ref{purepara}). The
left and right moving pure spinors $\l^{\a\a'}$ and $\bar
\l^{\ah\ah'}$ are five-dimensional symplectic Majorana spinors
(with a corresponding $SO(1,4)$ spinor index $\a$ or $\a'$) and
have an extra index $\a'$ and $\ah'$ in the spinor representation
of $SO(3)$.\footnote{The original six-dimensional pure spinors
$\l^{\a i}$ of (\ref{purity6}) is in the Weyl representation of
$SO(6)$ and has an additional index $i=1,2$ transforming as a
doublet of $SU(2)$. It decomposes naturally according to the local
symmetry group of our supercoset. The Weyl representation of
$SO(6)$ corresponds to the symplectic Majorana representation of
$SO(1,4)\simeq Sp(4)$, while the extra harmonic index $i$
corresponds precisely to the spinor index $\a'$ of $SO(3)$.} The
six-dimensional pure spinor constraint (\ref{purity6}), rewritten in
terms of the supercoset (\ref{supercoset5}), reads
 \bea
 C'_{\a'\b'}\l^{\a\a'}(C\gamma^m)_{\a\b}\l^{\b\b'}=0,&\qquad
 m=0,\ldots,4,\\
 C'_{\a'\b'}C_{\a\b}\l^{\a\a'}\l^{\b\b'}=0,&\nn
 \eea
where $\gamma^m$ and $C$ are the $SO(1,4)$ gamma matrices and
charge conjugation matrix and $C'$ is the $SO(3)$ charge
conjugation matrix. The pure spinor action is again BRST
invariant. As for the $\kappa$-symmetry discussed before, it seems
that in lower-dimensional pure spinor superstrings, the
BRST symmetry is related to the one-loop conformal invariance but
not to the Weyl invariance.

{\it Adding open strings}

The pure spinor sigma-model we have constructed above, even if it is
gauge invariant and BRST invariant at all order in perturbation
theory, does not correspond to a consistent non-critical
superstring. It is only consistent after adding an open string
sector. In particular, \cite{Klebanov:2004ya} suggested introducing
boundary conditions corresponding to uncharged space-filling
D-branes. We need uncharged D-branes because we do not want to
introduce any additional RR flux. They can be thought of as
space-filling brane-antibrane pairs.

When we put simultaneously D-brane and anti-D-brane boundary
conditions in a flat background, two things usually happen. This
completely breaks spacetime supersymmetry, since they preserve two
different sets of supercharges. An open string tachyon appears in the
spectrum. However, it was argued that, in the case of space-filling
branes and anti-branes on $AdS_5\times S^1$, the physics is different
from the flat space one. In particular, the space-filling brane
anti-brane system will break only half of the sixteen spacetime
supersymmetries.\footnote{In the six-dimensional linear dilaton
background, the system of $N_f$ space-filling brane/antibrane pairs,
together with $N_c$ D3 branes extending in the flat Minkowski part of
the space was studied in \cite{Fotopoulos:2005cn} and later inn
\cite{Murthy:2006xt}. For a finite number of colors and flavors, the
system preserves four supersymmetries, even if branes and anti-branes
are present simultaneously. We can regard the $AdS_5\times S^1$
background as the near horizon limit of the D3 branes of the linear
dilaton system, when the number of colors and flavors becomes very
large. As in the usual $AdS_5\times S^5$ case, in the near horizon
limit we double the number of supersymmetries, which gives a total of
eight supercharges, the appropriate number to match the dual
four-dimensional ${\cal N}=1$ SQCD in the conformal window.} We
suggest that this system will break the global $SU(2)$ symmetry of the
supercoset as well, leaving just a bosonic $SO(2,4)\times U(1)$ global
symmetry.  Moreover, the mass squared of the open string tachyon,
albeit negative, will be above the BF bound and therefore lead to no
instabilities. It would be interesting to prove these two conjectures,
by studying the spectrum of the worldsheet theory we have just
described. We leave this for future investigations.

It was suggested in \cite{Klebanov:2004ya} that the gauge theory dual
to this closed plus open superstring theory be four-dimensional ${\cal
N}=1$ SQCD at an IR superconformal fixed point. Note that holography
usually relates a closed superstring theory to a gauge theory, while
here we are considering a closed plus open superstring theory. The
$N_f$ brane anti-brane pairs correspond to the gauge theory
flavors. Provided that the two conjectures we discussed above are
indeed verified, the global symmetries on the two sides of the duality
are matched. The string theory global symmetries are $SO(2,4)\times
U(1)_R$, coming from the $AdS_5\times S^1$ isometries, and an
additional $SU(N_f)\times SU(N_f)$ flavor symmetry group that rotates
the space-filling branes and anti-branes. This fits nicely with the
global symmetry group of SQCD at the IR fixed point.

The first step in establishing this holographic duality would be
to compute the mass of the open string tachyon, which will depend
on the RR five-form flux $N_c$ and the number of flavor branes
$N_f$. One expects that it satisfies the BF bound when $N_f$ and
$N_c$ are inside the conformal window of the dual gauge theory.

\subsection{Ten-dimensional $AdS_p\times S^p\times{\cal M}$}

All the backgrounds of the kind $AdS_p\times S^p$ with RR $p$-form
flux correspond to the noncompact part of a ten-dimensional
background.\footnote{They have been studied in the hybrid formalism in
\cite{Berkovits:1999zq} for $p=2$ and in \cite{Berkovits:1999im} for
$p=3$. The matter part of the hybrid action is the same as the matter
part of the pure spinor action.  However their ghost sector is
different.} This is because the scalar curvature of these spaces
vanishes so they satisfy the one-loop beta-function equations
with $D=10$.\footnote{This agrees with the results of
\cite{Kuperstein:2004yk}. Those authors found that there is no
solution to the leading order non-critical supergravity equations for
these backgrounds when supported only by RR flux.} The case $p=5$
corresponds to the well known $AdS_5\times S^5$ background
\cite{Metsaev:1998it}. The lower-dimensional cases $p=2,3$, are
suitable for a Calabi-Yau compactification on a three- and a two-fold,
respectively.

{\it Curvature equation}

In \cite{Berkovits:1999zq} it was shown that if $G$ is a Ricci
flat supergroup, then the scalar curvature of the supercoset $G/H$
is equal to the curvature of its bosonic subgroup, namely the
bosonic $AdS_p\times S^p$ manifold. The $AdS$ and the sphere give
the same contribution but with opposite sign\footnote{Note that
in the supercoset construction the radii of $AdS_p$ and $S^p$ need
to be equal in order to preserve the superalgebra.}, so the total
scalar curvature of the supercoset vanishes. We will review now how
this cancellation works at the level of the super Ricci curvature
itself.

We denote by $a,b=1,\ldots,p$ the vector indices along $AdS_p$ and
by $a',b'=1,\ldots,p$ the vector indices along the $S^p$. The
bosonic legs of the super Ricci curvature are
 \bea
 R_{ab}&=&\frac{1}{16}\{\g_{a},\g_{b}\}^{\a}{}_{\a}\d^{\a'}{}_{\a'}-\eta_{as}\d^{c}{}_{c}+\eta_{ab}\\
 &=&(\frac{1}{8}S_{AdS_{p}}S_{S^{p}}-V_{AdS_{p}}+1)\eta_{ab},\nn\\
 R_{a'b'}&=&-\frac{1}{16}
 \{\g_{a'},\g_{b'}\}^{\a'}{}_{\a'}\d^{\a}{}_{\a}-\eta_{a'b'}\d^{c'}{}_{c'}+\eta_{a'b'}\\
 &=&-(\frac{1}{8}S_{S^{p}}S_{AdS_{p}}-V_{S_{p}}+1)\eta_{a'b'}
 \eea
where $\eta_{ab}$ and $\eta_{a'b'}$ are the bosonic metrics on the
supergroup and $\a$ are the spinor indices on $AdS_p$ while $\a'$
are the spinor indices on $S^p$. $S_{AdS_{p}}$ and $S_{S_{p}}$
stand for the dimension of the spinor representation of the
$AdS_{p}$ and $S^p$ part of the super-algebra, similarly
$V_{AdS_{p}}$ and $V_{S_{p}}$ are the bosonic dimensions of the
$AdS_p$ and $S^{p}$ spaces. It is clear that the bosonic $AdS_p$
and the $S^p$ contributions to the scalar curvature cancel each
other. For the Fermionic part of the super Ricci curvature we have
\begin{eqnarray}
 R(G/H)_{\a\a'\bh\bh'} & = & \frac{1}{8} \Big[ - C'_{\a'\bh'}
 (C\g^{a}\g_{a})_{\a\bh} + C_{\a'\bh'}(C'\g^{a'}\g_{a'})_{\a\bh} -
 C'_{\bh'\a'}(C\g^{a}\g_{a})_{\bh\a} + \nonumber \\
 && {} + C_{\bh'\a'}(C'\g^{a'}\g_{a'})_{\bh\a} \Big]
 - \frac{1}{2}\left(C'_{\a'\bh'}(C\g^{cd}\g_{cd})_{\a\bh} +
 C_{\a'\bh'}(C'\g^{c'd'}\g_{c'd'})_{\a\bh}\right) = \nonumber \\
 & = & \frac{1}{2}\Big[(V_{S^{p}})^2-(V_{AdS_{p}})^2\Big]C'_{\a'\bh'}C_{\a\bh}
 \ ,
\end{eqnarray}
where $C$ and $C\,'$ are the charge conjugation matrices of the
$AdS_p$ and $S^p$ part of the supergroup. We find that the
fermionic part of the super Ricci curvature vanishes identically
 \bea
 R(G/H)_{\a\a'\bh\bh'}=0.
 \eea
As a result, the supertrace of the super Ricci curvature vanishes.

\subsection{Central charge}

In order to verify that our models are consistent string theories we
need to see that their total central charge vanishes.
Since the sigma-models are strongly coupled, it is hard to compute
the exact central charge at their fixed point. What we can do is
to consider the ``naive" central charge that one would get in the
small curvature limit, i.e., in the classical sigma-model. In the
pure spinor sigma-model, we just add up the contribution to the
central charge coming from each CFT separately, since in the small
curvature limit they are just free and decoupled. We are in fact
just computing the flat space central charge. In each sector, the
matter part is given by the bosons $\{X^a\}$, for $a=1,\ldots,2d$,
which are worldsheet scalars, and the supercoordinates and their
conjugate momenta $\{p_\a,\t^\a\}$, which have weight $(1,0)$,
while the pure spinor beta-gamma system $\{w_\a,\l^\a\}$ has
weight $(1,0)$ and has been described in section
\ref{puresection}. It turns out that in all different dimensions,
the matter central charge is exactly cancelled by the ghost
central charge. The field content is the same for both the
non-critical and the critical models (see \cite{Adam:2006bt} and
\cite{Grassi:2005sb,Wyllard:2005fh}). We summarize the various
contribution to the vanishing central charge in different
dimensions as follows
 \be
 \begin{array}{ccccccc}
 d=2: &\quad& c_{tot}=&(2)_{\{X\}}+ &(-4)_{\{p,\t\}}&+(2)_{\{w,\l\}}&=0,\\
 d=4: &\quad& c_{tot}=&(4)_{\{X\}}+ &(-8)_{\{p,\t\}}&+(4)_{\{w,\l\}}&=0,\\
 d=6: &\quad& c_{tot}=&(6)_{\{X\}}+ &(-16)_{\{p,\t\}}&+(10)_{\{w,\l\}}&=0.
 \end{array}
 \ee
This counting of degrees of freedom is of course heuristic. The
evaluation of the exact central charge requires to solve for the
spectrum of the model. One way of doing it is by making use of the
Bethe ansatz approach developed in \cite{Mann:2005ab}. We hope to
report about this in the future.

\section{Discussion and open problems}

In this paper we have constructed the worldsheet theory of type II
superstrings on $AdS$ backgrounds with RR flux, which are realized
as supercosets $G/H$ where $G$ has a $\ZZ_4$ automorphism. We
have shown in particular that in all such backgrounds string
theory is quantum integrable. This holds both for the non-critical
and for the ten-dimensional superstrings, in particular for the
topological sector of the latter. A nice feature we
found\footnote{This was also noticed in \cite{Chen:2005uj} for the
GS sigma models.} is that the dependence of the Lax connection
$a(\mu)$ on the spectral parameter
(\ref{eq:AdS-flat-current-params}) is the same in all models. Once
we established the existence of these type II backgrounds, there
are many directions that open up for a future investigation. Let
us list some of them.

The worldsheet theory for non-critical strings is a strongly coupled
sigma-model, whose coupling is given by the curvature of $AdS$. Due to
the lack of tools to analyze strongly coupled sigma-models we could
prove neither exact conformal invariance nor Weyl invariance. It would
be interesting to prove the existence of the fixed point
non-perturbatively, or at least to all orders in perturbation
theory.\footnote{For the backgrounds of the kind $AdS_p\times S^p$,
the beta-function vanishes at all orders in perturbation theory
\cite{Berkovits:1999im}. However, the method used in those cases
relies on an extension of the supergroup $G$ which is not possible in
our non-critical cases.} In the case of $AdS_2$ more can be said. The
proofs of gauge invariance, BRST invariance and integrability of the
sigma-model that we discussed above hold in general to all orders in
the worldsheet perturbation theory.  The non-perturbative
contributions to the action, on the other hand, come in the form of
worldsheet instantons and are counted by the factor $e^{-{1\over
\l^2}}$, where $\l$ is the sigma-model coupling. In the $AdS_2$ case
there are no two-cycles the worldsheet instantons can wrap on. Hence,
the sigma-model does not receive any non-perturbative corrections and
it is gauge invariant, BRST invariant and integrable exactly.

Once it is established that these superstring sigma-models are
integrable, it is natural to look for their spectrum. The computation
of the quantum spectrum of string theory on $AdS_5\times S^5$ based on
the $PSU(2,2\vert 4)$ supercoset is a formidably hard task. Our lower
dimensional non-critical string theories might be easier to solve
since they are described by somewhat simpler supercosets. The $AdS_2$
background is probably the simplest example of a type II RR
background and we have argued that the sigma-model is exact
non-perturbatively, due to the absence of worldsheet instantons. Since
the spacetime is two-dimensional, the semiclassical spectrum is
empty. The next non-trivial example is the $AdS_4$ background, for
which the semiclassical spectrum contains for example spinning string
solutions. As a first step, it would be interesting to work out the
complete classical spectrum, encoded in the algebraic curve method of
\cite{Beisert:2005bm}, which fully exploits the integrability
properties of the classical sigma-model. In order to look for the
spectrum of the quantum sigma-model, one can follow two different
approaches. A first way is by computing the pure spinor
cohomology. The second approach makes use of the Bethe ansatz, as
proposed in \cite{Mann:2005ab}. In the latter paper, the authors
focussed on a toy model, based on the supercoset $Osp(2m+2\vert
2m)$. It seems plausible that our sigma-model for $AdS_4$, which we
realized as a $Osp(2\vert 4)$ supercoset, might be solvable in the
same spirit. The exact solution will fix also the value of the central
charge at the strongly coupled fixed point, which we have not been able
to evaluate.

Another issue pertains to the interpretation of the
ten-dimensional backgrounds $AdS_p\times S^p\times CY_{5-p}$,
whose non-compact part we have discussed in detail. While their
GS formulation certainly describes the full compactified
superstring, the interpretation of the pure spinor formulation is
still not completely clear. Recently, there have been different
proposals regarding the pure spinor superstring compactified on
Calabi-Yau
\cite{Berkovits:2005bt,Grassi:2005sb,Wyllard:2005fh,Chandia:2005fi,Grassi:2005jz}.
In the case in which the background is flat four-dimensional
Minkowski times a CY three-fold, it has been argued that the pure
spinor formalism computes only the topological amplitudes of the
full superstring \cite{Berkovits:2005bt}. It would be interesting
to understand what happens in our backgrounds.

In section \ref{secD} we considered the addition of boundary
conditions to the sigma-model, in particular space-filling
D-branes. The classification of branes in non-compact spaces can be in
general a hard task, even when they are supported only by NS-NS
flux. Since very little is known about such classification on RR
backgrounds, it would be nice to make progress in this analysis. In
particular, the pure spinor formalism seems a convenient starting
point for such a search, due to his simple couplings to RR
backgrounds.

An interesting open problem is to figure out how holography works
for the non-critical backgrounds and to identify the field theory
duals to these new non-critical $AdS_{2d}$ backgrounds if they
are field theories at all. In particular, the existence of the
infinite set of nonlocal charges on the string sigma-model is
related to the Yangian symmetry of the dual gauge theory. In the
$AdS_5\times S^5$ case, the existence of the Yangian symmetry has
been established in the dual ${\cal N}=4$ gauge theory at weak
coupling \cite{Dolan:2003uh}. It would be nice to identify the
Yangian symmetry on the gauge dual side in all our cases.

The duals of the non-critical superstrings on $AdS_{2d}$ are in
general strongly coupled $2d-1$ superconformal field theories. A
particularly interesting example would be the relation between the
type IIB non-critical superstring on $AdS_5\times S^1$ with
space-filling branes and four-dimensional ${\cal N}=1$ SQCD, which
was suggested in \cite{Klebanov:2004ya} by looking at the
six-dimensional non-critical supergravity. In this case there is
no decoupling limit, namely the string dual of the conformal
window of SQCD should contain closed as well as open strings. It
would be nice to analyze our worldsheet theory for such
background. Firstly, one should check the vanishing of the
one-loop beta-function in the presence of boundaries. Only with
the contribution coming from the boundary of the worldsheet should
the theory be Weyl invariant at one-loop. This is equivalent to
the statement, reviewed in Appendix \ref{sixsugra}, that the
non-critical supergravity equations of motion for both the metric
and the dilaton are satisfied only with the inclusion of the
space-filling branes. Then, one should show that, at least
perturbatively in the sigma-model coupling, the mass of the open
string tachyon lies above the BF bound. An interesting thing to
study is T-duality along the circle, which might be related to
Seiberg duality in the dual SQCD \cite{Klebanov:2004ya}.

Finally, it would be interesting to see what happens to the Yangian
symmetry of the closed string sector once we add boundaries. The Bethe
ansatz for open strings on $AdS_5 \times S^5$ was discussed in
\cite{McLoughlin:2005gj} and in a recent paper \cite{Mann:2006rh} it
has been shown that for certain choices of boundary conditions the
bosonic part of the open string sector is still integrable. It would
be interesting to check whether this is true for the $AdS_5\times S^1$
model with space-filling branes. In the case in which the open string
sector is integrable, it would be intriguing to investigate the
implications of such a symmetry in the dual SQCD.

\acknowledgments We would like to thank A.~Babichenko, N.~Berkovits,
O.~Chandia, P.A.~Grassi, N.~Itzhaki, J.~Polchinski and S.~Yankielowicz
for useful discussions. L.M.\ would like to thank Cobi Sonnenschein
for sharing some unpublished results on non-critical supergravity.
I.A.\ and L.M.\ would like to thank the organizers of the S\~ao Paulo
Workshop on Pure Spinors in Superstring Theory for the stimulating
environment, where part of this work was done.

\appendix

\section{$\kappa$-symmetry and torsion constraints} \label{sec:kappa-symm-and-torsion}

In section \ref{gssm} we constructed the Green-Schwarz action on a
supergroup with $\ZZ_4$ automorphism and we found that the
structure constants of the supergroup have to obey the relation
(\ref{eq:kappa-symmetry-condition}) rewritten here
\begin{equation}\label{relek}
   \eta_{\beta \hat \beta} \left( f_{a \hat \alpha}^\beta
  f_{b \alpha}^{\hat \beta} + f_{b \hat \alpha}^\beta f_{i
  \alpha}^{\hat \beta} \right) = c_{\alpha \hat \alpha} \eta_{a
  b},
\end{equation}
where $c_{\a\ah}$ is some symmetric matrix, in order for the
action to be $\kappa$-symmetric. We will show now how this
relation is equivalent to the torsion constraints of the
supergravity background.

Recalling that the spacetime torsion is defined as
 \bea
 [\nabla_A,\nabla_B\}&=&T^C_{AB}\nabla_C,
 \eea
we find that the structure constants $f_{a \hat \alpha}^\beta$ and
$f_{b \alpha}^{\hat \beta}$ correspond to some particular
components of the torsion
 \bea
 f_{a\alpha}^{\bh}=-T_{a\alpha}^\bh,&f_{a\ah}^{\b}=-T_{a\ah}^\b.
 \eea
Let us specialize to the case $AdS_p\times S^p$ first, namely the
type II superstring compactified on a Calabi Yau. In type II
supergravity the torsion is related to the RR superfield
$P^{\b\bh}$ by the following constraint \cite{Berkovits:1995cb}
 \bea
 T_{a\a}^\bh=(\gamma_a)_{\a\b}P^{\b\bh},&T_{a\ah}^\b=(\gamma_a)_{\ah\bh}P^{\b\bh},
 \label{torsioncon}
 \eea
and the RR superfield is given by
 \bea
 P^{\b\bh}={g_S\over p!}(\gamma_{m_1\ldots m_p})^{\b\bh}F^{m_1\ldots
 m_p}=g_S N_c\delta^{\b\bh}
 \eea
where $F_p$ is the self dual $p$-form flux and the $\gamma_a$ are
the Pauli matrices, namely the off diagonal blocks of the Dirac
matrices of $SO(1,p-1)$. The structure constants are given in
Appendix \ref{adspsp}As a last ingredient, we notice that the
metric on the supergroup is proportional to the inverse of the RR
superfield $\eta_{\b\bh}\propto(P^{-1})_{\b\bh}\propto
\delta_{\b\bh}$. Putting everything together, we can cast the
relation (\ref{relek}) in the form
 \bea
 \eta_{\beta \hat \beta} \left( f_{a \hat \alpha}^\beta
  f_{b \alpha}^{\hat \beta} + f_{b \hat \alpha}^\beta f_{i
  \alpha}^{\hat \beta}
  \right)=\{\gamma_a,\gamma_b\}_\a^\b\delta_{\b\ah}=2\eta_{ab}\delta_{\a\ah},
 \eea
so we find that the symmetric matrix $c_{\a\ah}=\delta_{\a\ah}$ is
proportional to the inverse RR flux.

In the case of the non-critical $AdS_2$, $AdS_4$ and $AdS_5\times
S^1$, the same result follows, provided an analogous torsion
constraint (\ref{torsioncon}) is imposed. This can be understood again
as a supergravity constraint of ${\cal N}=(2,2)$ in two dimensions
\cite{Berkovits:2001tg} or ${\cal N}=2$ in four and six dimensions
\cite{Berkovits:1995cb}.

\section{Pure spinor sigma-models}

In this appendix we collect some computations used in the main
text for the pure spinor superstrings. We refer to sections 2 and
3 for the notations.

\subsection{The pure spinor sigma-model from BRST symmetry}
\label{sec:pure-spinor-sigma-model-from-BRST}

Using the fact that $\langle A B \rangle \neq 0$ only for $A \in
\mathcal{H}_r$ and $B \in \mathcal{H}_{4 - r}$, $r = 0, \dots, 3$
\cite{Berkovits:1999zq}, the most general matter part which has a
global symmetry under left multiplication by elements of $G$ and
is invariant under the gauge symmetry $g \simeq g h$, where $h \in
H$, is
\begin{displaymath}
  \int d^2 z \langle \alpha J_2 \bar J_2 + \beta J_1 \bar J_3 + \gamma
  J_3 \bar J_1 + \delta J_3 \bar d + \epsilon \bar J_1 d - f d \bar d
  \rangle \ ,
\end{displaymath}
where we used the Lie-algebra valued field $d$, $\bar d$ defined
by $d = d_\alpha \eta^{\alpha \hat \alpha} T_{\hat \alpha}$, $\bar
d = \bar d_{\hat \alpha} \eta^{\alpha \hat \alpha} T_\alpha$ and
$f$ is the RR-flux. While in flat background the $d$'s are
composite fields, in curved backgrounds they can be treated as
independent fields.

The pure spinor part includes the kinetic terms $\langle w \bar
\partial \lambda \rangle$ and $\langle \bar w \partial \bar \lambda
\rangle$ for the pure spinor $\beta \gamma$-systems. Since these
terms are not gauge invariant, they must be accompanied by terms
coupling the pure spinor gauge generators with the matter gauge
currents $\langle N \bar J_0 + \bar N J_0 \rangle$ in order to
compensate. The backgrounds we are considering also require
additional terms which must be gauge invariant under the pure
spinor gauge transformation of $w$ and $\bar w$ and hence must be
expressed in terms of the Lorentz currents and the ghost currents
$J_\mathrm{gh} = \langle w \lambda \rangle$ and $\bar
J_\mathrm{gh} = \langle \bar w \bar \lambda \rangle$ (such terms
are given by $S_{\alpha \hat \gamma}^{\beta \hat \delta}$ in the
Type II action in \cite{Berkovits:2001ue}). The additional term
required is $\langle N \bar N \rangle$.

Therefore the sigma-model is of the form
\begin{eqnarray}
  S & = & \int d^2 z \langle \alpha J_2 \bar J_2 + \beta J_1 \bar J_3 +
  \gamma J_3 \bar J_1 + \delta J_3 \bar d + \epsilon \bar J_1 d - f d
  \bar d + \nonumber \\
  && {} + w \bar \partial \lambda + \bar w \partial \bar \lambda + N
  \bar J_0 + \bar N J_0 + a N \bar N \rangle
\end{eqnarray}
and the accompanying BRST-like operator is
\begin{equation} \label{eq:unintegrated-BRST}
  Q_B = \oint \langle dz \lambda d - d\bar z \bar \lambda \bar d
  \rangle \ .
\end{equation}
By integrating out $d$ and $\bar d$ and redefining $\gamma \to
\gamma + \frac{\epsilon \delta}{f}$ one gets
\begin{equation}
  S = \int d^2 z \langle \alpha J_2 \bar J_2 + \beta J_1 \bar J_3 +
  \gamma J_3 \bar J_1 + w \bar \partial \lambda + \bar w \partial \bar
  \lambda + N \bar J_0 + \bar N J_0 + a N \bar N \rangle \ .
  \label{action1}
\end{equation}
After rescaling $\lambda \to \frac{\delta}{f} \lambda$, $w \to
\frac{f}c{\delta} w$, $\bar \lambda \to \frac{\epsilon}{f} \bar
\lambda$, $\bar w \to \frac{f}{\epsilon} \bar w$ the BRST currents
are $j_B = \langle \lambda d \rangle = \langle \lambda J_3
\rangle$ and $\bar j_B = \langle \bar \lambda \bar d \rangle =
\langle \bar \lambda \bar J_1 \rangle$. The BRST charge
(\ref{eq:unintegrated-BRST}) now reads
\begin{equation}
  Q_B = \oint \langle dz \lambda J_3 + d\bar z \bar \lambda \bar J_1
  \rangle \ .
\end{equation}

The coefficients of the various terms will be determined by
requiring the action to be BRST invariant, \textit{i.e.}\ the BRST
currents are holomorphic and the corresponding charge is
nilpotent.

From the action (\ref{action1}) we derive the following equations
of motion
 \begin{eqnarray}
  (\beta + \gamma) \bar \nabla J_3 & = & (2 \beta - \alpha) [J_1,
  \bar J_2] + (\alpha + \beta - \gamma) [J_2, \bar J_1] + [N, \bar
  J_3] + [\bar N, J_3] \ , \\
  (\beta + \gamma) \nabla \bar J_1 & = & (\alpha - 2 \beta) [J_2, \bar
  J_3] + (\gamma - \alpha - \beta) [J_3, \bar J_2] + [N, \bar J_1] +
  [\bar N, J_1] \ , \\
  \bar \nabla \lambda & = & - a [\bar N, \lambda] \ , \quad
  \nabla \bar \lambda  = - a [N, \bar \lambda] \ .
\end{eqnarray}
After one takes into account that $[N, \lambda] = 0$ because of
the pure spinor condition $\{ \lambda, \lambda \} = 0$
\cite{Berkovits:2004xu}, requiring $\bar \partial j_B = 0$  leads
to the equations
\begin{equation}
  \beta + \gamma = 1 \ , \quad \alpha = 2 \beta \ , \quad \alpha +
  \beta = \gamma \ , \quad a = -1 \ ,
\end{equation}
whose solution is
\begin{equation}
  \alpha = \frac{1}{2} \ , \quad \beta = \frac{1}{4} \ , \quad
  \gamma  = \frac{3}{4} \ , a = -1 \ .
\end{equation}
With this solution it is easy to check that
\begin{equation}
  \partial \bar j_B = \langle [\bar \lambda, \bar N] J_1 \rangle \ ,
\end{equation}
which again vanishes because of the constraint $\{ \bar \lambda,
\bar \lambda \} = 0$. The proof of the nilpotence of the BRST
charge then follows just as in \cite{Berkovits:2004xu}.

Hence the pure spinor sigma-model is
\begin{equation}
  S = \int d^2 z \left\langle \frac{1}{2} J_2 \bar J_2 + \frac{1}{4} J_1
  \bar J_3 + \frac{3}{4} J_3 \bar J_1 + w \bar\partial  \lambda + \bar w
  \partial \bar \lambda + N \bar J_0 + \bar N J_0 -N \bar N
  \right\rangle
\end{equation}
for all dimensions and this of course matches the critical case as
well.

\subsection{BRST invariance of the conserved pure spinor charges}

\label{sec:ps-charges-BRST-inv} We would like to verify that the
conserved charges (\ref{eq:ps-conserved-charges}) are BRST
invariant. This requirement stems from the fact that for a charge
to be a symmetry, it is not sufficient for it to be conserved. A
symmetry maps physical states (states in the pure spinor
cohomology) to other physical states. For this to happen, the
charge itself must be BRST-closed.

The BRST transformations of the various worldsheet fields are
given by
\begin{eqnarray}
  \delta_B g & = & g (\epsilon \lambda + \epsilon \bar \lambda) \ ,
  \quad  \delta_B w = -J_3 \epsilon \ , \quad \delta_B \bar w = -\bar
  J_1 \epsilon \ , \quad \delta_B \lambda = \delta_B \bar \lambda = 0 \ , \\
  \delta_B J_0 & = & [J_3, \epsilon \lambda] + [J_1, \epsilon \bar
  \lambda] \ , \\
  \delta_B J_1 & = & \partial (\epsilon \lambda) + [J_0, \epsilon
  \lambda] + [J_2, \epsilon \bar \lambda] \ , \\
  \delta_B J_2 & = & [J_1, \epsilon \lambda] + [J_3, \epsilon \bar
  \lambda] \ , \\
  \delta_B J_3 & = & \partial (\epsilon \bar \lambda) + [J_2, \epsilon
  \lambda] + [J_0, \epsilon \bar \lambda] \ , \\
  \delta_B N & = & \{ J_3 \epsilon, \lambda \} \ , \quad
  \delta_B \bar N = \{ \bar J_1 \epsilon, \bar \lambda \} \ .
\end{eqnarray}

In order to demonstrate that the charges
(\ref{eq:ps-conserved-charges}) are indeed BRST closed, we define
the following operator
\begin{equation}
  U(x, \bar x; y, \bar y) = \mathrm{P} \exp \left[ - \int_y^x \left(
  dz a + d\bar z \bar{a} \right) \right] \ .
\end{equation}
The BRST variation of $U_C$ can now be written as
\begin{eqnarray} \label{eq:AdS-charge-BRST-var}
  \delta_B U_C &  = & - \int_C dz U(x, \bar x; z, \bar z) \delta_B a
  (z, \bar z) U(z, \bar z; y, \bar y) - \nonumber \\
  && {} - \int_C d\bar z U(x, \bar x; z, \bar z) \delta_B \bar{a}
  (z, \bar z) U(z, \bar z; y, \bar y) \ ,
\end{eqnarray}
where
\begin{eqnarray}
  \delta_B a & = & g \Big[ c_2 ([J_1, \epsilon \lambda] +
  [J_3, \epsilon \bar \lambda]) + c_1 (\epsilon \partial \lambda + [J_0,
  \epsilon \lambda] + [J_2, \epsilon \bar \lambda]) + \nonumber \\
  && {} + c_3 (\epsilon \partial \bar \lambda + [J_2, \epsilon \lambda] + [J_0,
  \epsilon \bar \lambda]) + c_N \{ J_3 \epsilon, \lambda \} + [\epsilon
  \lambda + \epsilon \bar \lambda, A] \Big] g^{-1} \ ,\\
  \delta_B \bar{a} & = & g \Big[ \bar c_2 ([\bar J_1, \epsilon
  \lambda] + [\bar J_3, \epsilon \bar \lambda]) + \bar c_1 (\epsilon \bar
  \partial \lambda + [\bar J_0, \epsilon \lambda] + [\bar J_2,
  \epsilon \bar \lambda]) + \nonumber \\
  && {} + \bar c_3 (\epsilon \bar \partial \bar \lambda + [\bar J_2, \epsilon
  \lambda] + [\bar J_0, \epsilon \bar \lambda]) + \bar c_N \{ \bar J_1
  \epsilon, \bar \lambda \} + [\epsilon \lambda + \epsilon \bar
  \lambda, \bar A] \Big] g^{-1}
\end{eqnarray}
and $g: \Sigma \to G$ is the mapping from the worldsheet to the
supergroup $G$.

The derivative terms in the first integral in
(\ref{eq:AdS-charge-BRST-var}) can be computed by integrating by
parts, so the derivative terms turn out to be
\begin{displaymath}
  \int_C dz U(x, \bar x; z, \bar z) g(z, \bar z) \left[ [c_1
  \epsilon \lambda + c_3 \epsilon \bar \lambda, A] + [c_1 \epsilon
  \lambda + c_3 \epsilon \bar \lambda, J] \right]_{z, \bar z}
  g(z, \bar z)^{-1} U(z, \bar z; y, \bar y) \ .
\end{displaymath}
Plugging this back into the integral and collecting all terms one
gets
\begin{eqnarray}
  I & = & \int_C dz U(x, \bar x; z, \bar z) g(z, \bar z) \Big[
  (c_2 - c_1 (c_1 + 2)) [J_1, \epsilon \lambda] + (c_3 - c_2 (c_1 + 1)
  -c_1) [J_2, \epsilon \lambda] + \nonumber \\
  && {} + (c_N - c_3 (c_1 + 1) - c_1) [J_3, \epsilon \lambda] + (c_2 -
  c_3 (c_3 + 2) [J_3, \epsilon \bar \lambda]  + c_N (c_1 + 1)
  [\epsilon \lambda, N] + \nonumber \\
  && {} + (c_1 - c_2 (c_3 + 1) -c) [J_2, \epsilon \bar \lambda] + (c_1
  (c_3 + 1) + c) [\epsilon \bar \lambda, J_1] + \nonumber \\
  && {}  + c_N (c_3 + 1) [\epsilon \bar
  \lambda, N] \Big]_{z, \bar z} g(z, \bar z)^{-1} U(z, \bar z;
  y, \bar y) \ .
\end{eqnarray}
Note that $[\lambda, N] = 0$ by using the Jacobi identity and the
pure spinor constraint $\{ \lambda, \lambda \} = 0$. The last term
is handled by using the equation of motion $\nabla \bar \lambda =
[N, \bar \lambda]$
\begin{eqnarray*}
  \lefteqn{\int_C dz U(x, \bar x; z, \bar z) g(z, \bar z)
  \left. [\epsilon \bar \lambda, N]
  \right|_{z, \bar z} g(z, \bar z)^{-1} U(z, \bar z; y, \bar y)
  =}\\
  & = &  - \int_C dz U(x, \bar x; z, \bar z) g(z, \bar z) \Big[
  [\epsilon \bar \lambda, A] + [\epsilon \bar \lambda, J_1 + J_2 +
  J_3] \Big]_{z, \bar z} g(z, \bar z)^{-1} U(z, \bar z; y, \bar
  y) \ .
\end{eqnarray*}
Using this for a fraction $x$ of $[\epsilon \bar \lambda, N]$
yields that the integral evaluates to
\begin{eqnarray}
  I & = & \int_C dz U(x, \bar x; z, \bar z) g(z, \bar z)
  \Big[ (c_2 - c_1 (c_1 + 2)) [J_1, \epsilon \lambda] + (c_3 - c_2
  (c_1 + 1) -c_1) [J_2, \epsilon \lambda] + \nonumber \\
  && {} + (c_N - c_3 (c_1 + 1) -c_1) [J_3, \epsilon \lambda] +  (c_2 -
  c_3 (c_3 + 2) + x (c_3 + 1)) [J_3, \epsilon \bar \lambda] +
  \nonumber \\
  && {} + (c_1 - c_2 (c_3 + 1) - c_3 + x (c_2 + 1)) [J_2, \epsilon
  \bar \lambda] + (c_1 (c_3 + 1) + c_3 - x (c_1 + 1)) [\epsilon \bar
  \lambda, J_1] + \nonumber\\
  && {} + (c_N (c_3 + 1) - x (c_N + 1)) [\epsilon \bar \lambda, N] \Big]_{z,
  \bar z} g(z, \bar z)^{-1} U(z, \bar z; y, \bar y) \ .
\end{eqnarray}
After choosing $x = \frac{c_N (c_3 + 1)}{c_N + 1}$ and
substituting (\ref{eq:AdS-flat-current-params}) we get $I=0$. The
second integral in (\ref{eq:AdS-charge-BRST-var}) vanishes in a
similar way, so the conserved charges found here are indeed BRST
invariant.

\subsection{Ghost number one cohomology} \label{empti}

In this appendix we will prove the claim made in section
\ref{quantumbrst} that the classical BRST cohomology of integrated
vertex operators $\int d^2z \langle {\cal O}^{(1)}_{z\bar
z}\rangle$ at ghost number one is empty.

The most general ghost number one gauge-invariant integrated vertex operator
is
 \bea
 \langle {\cal O}_{z\bar z}^{(1)} \rangle&=&
 \langle a_1\bar J_2[J_3,\e\bar\l]+\bar a_1 J_2[\bar
 J_1,\e\l]+a_2\bar J_2[J_1,\e\l]+\bar a_2 J_2[\bar J_3,\bar
 \l]\nn\\
 &&+a_3J_3[\bar N,\e\l]+\bar a_3\bar
 J_1[N,\e\bar\l]+a_4J_3\bar\nabla(\e\l)+\bar a_4\bar
 J_1\nabla(\e\bar\l)\rangle,\label{ghostone}
 \eea
where we have written all the independent terms up to integrating
by parts on the Maurer-Cartan equations. We will consider the
insertion of a boundary at the end and concentrate on the bulk
terms first. The BRST variation of the  operator (\ref{ghostone})
consists of three different kind of terms
 \bea
 \e'Q\langle {\cal O}_{z\bar z}^{(1)}
 \rangle&=&\Omega_1+\Omega_2+\Omega_3+\textrm{e.o.m.'s}+\textrm{pure gauge},
 \eea
where we have omitted terms proportional to the ghost equations of
motion (\ref{eq:AdS-N-eq}) and to the gauge transformations
parameterized by $\{\l,\bar\l\}$. We have to impose that the three
terms $\Omega_i$ vanish separately. The first term is
 \bea
 \Omega_1=(a_3+a_4-\bar a_3-\bar a_4)\langle \bar
 \nabla(\e\l)\nabla(\e'\bar\l)\rangle,
 \eea
so we demand
 \bea\label{coe2}
 a_3+a_4&=&\bar a_3+\bar a_4.
 \eea
Imposing the vanishing of the second term
 \bea
 \Omega_2&=&\langle(a_1-\bar a_1+a_3+a_4-\bar a_3-\bar
 a_4)[J_3,\e\bar\l]+(a_2-\bar a_2)[\bar
 J_3,\e'\bar\l][J_1,\e\l]\nn\\
 &&+(a_1-\bar a_1+a_2-\bar
 a_2)[J_2,\e'\l][\bar J_2,\e\bar\l]\rangle,
\eea we find the additional conditions
 \bea\label{coe1}
 a_1=\bar a_1,&\quad& a_2=\bar a_2.
 \eea
Finally, the third term reads
 \bea
 \Omega_3&=&\langle(a_2+\bar a_1)[\bar J_1,\e'\l][J_1,\e\l]+ (a_1+\bar a_2)[J_3,\e\bar\l][\bar
 J_3,\e'\bar\l]\nn\\
 &&-a_4[ J_3,\e\l][\bar J_3,\e'\l]-\bar a_4[\bar
 J_1,\e\bar\l][J_1,\e'\bar\l]\rangle.\label{termini}
 \eea
If we expand on the supergroup generators, the first term on the
right hand side is proportional to $ \l^\a\l^\b
\langle[T_\delta,T_\a][T_\rho,T_\b]\rangle$, where we summarized
with a greek letter the various spinor properties of the
supercharges and the pure spinors in the various dimensions. In
all dimensions, due to the supersymmetry algebra, the term inside
the supertrace is proportional to
$(\sigma^m)_{\delta\a}(\sigma_m)_{\b\rho}$, where
$\sigma^m_{\a\b}$ are the off diagonal blocks of the Dirac
matrices. Now comes the crucial property of our lower-dimensional
pure spinors, which behave precisely like the ten-dimensional
ones. In dimension $d=2n$, the product of two pure spinors is
always proportional to the middle dimensional form
$\sigma^{m_1\ldots m_n}_{\a\b}$, therefore the terms in
(\ref{termini}) are all proportional to
 \bea
 \sigma^m\sigma^{m_1\ldots m_n}\sigma_m,
 \eea
but this expression vanishes in all even dimensions due to the
properties of the gamma matrix algebra. Thus, in all dimensions we
find that $\Omega_3=0$ identically. As a result, imposing that
$\int d^2z\langle{\cal O}_{z\bar z}^{(1)}\rangle$ is BRST closed
requires that the coefficients $a_i,\bar a_i$ satisfy (\ref{coe2})
and (\ref{coe1}).

On the other hand, the following operator
 \bea
 \Sigma^{(0)}_{z\bar z}&=&-a_2\bar J_2J_2+(a_1-a_2)\bar J_1 J_3+(a_3-\bar
 a_4+a_2-a_1)N\bar N\nn\\&&+(a_4+a_1-a_2)w\bar\nabla\l+(\bar
 a_4+a_1-a_2)\bar w\nabla\bar\l,
 \eea
is such that
 \bea
 Q\int d^2z\langle\Sigma^{(0)}_{z\bar z}\rangle=\int
 d^2z\langle{\cal O}^{(1)}_{z\bar z}\rangle,
 \eea
so the cohomology for integrated vertex operators at ghost number
one is empty.

\section{Supergroups}
\label{superapp}

In this Appendix we list the details of the superalgebras we need
to realize the various backgrounds in the text. We constructed our
superalgebras according to \cite{Frappat:1996pb}.

\subsection{Notations}
Our notations follow the ones used by \cite{Pais:1975hu}. The superalgebra
satisfies the following commutation relations:
 \be
 [T_{m},T_{n}]=f^{p}_{mn}T_{p}
 \ee
  \be
 [T_{m},Q_{\a}]=F^{\b}_{m\a}Q_{\b}
 \ee
 \be
 \{Q_{\a},Q_{\b}\}=A^{m}_{\a\b}T_{m}
 \ee
where the $T$'s are the bosonic (Grassman even) generators of a
Lie algebra and the $Q$'s are the fermionic (Grassman odd)
elements. The indices are $m=1,...,d$ and $\a=1,...,D$. The
generators satisfy the following super-Jacobi identities:
 \be
f^{p}_{nr}f^{q}_{mp}+f^{p}_{rm}f^{q}_{np}+f^{p}_{mn}f^{q}_{rp}=0
 \ee
 \be
F^{\g}_{n\a}F^{\d}_{m\g}-F^{\g}_{m\a}F^{\d}_{n\g}-f^{p}_{mn}F^{\d}_{p\a}=0
 \ee
 \be
F^{\d}_{m\g}A^{n}_{\b\d}+F^{\d}_{m\b}A^{n}_{\g\d}-f^{n}_{mp}A^{p}_{\b\g}=0
 \ee
 \be
A^{p}_{\b\g}F^{\d}_{p\a}+A^{p}_{\g\a}F^{\d}_{p\b}+A^{p}_{\a\b}F^{\d}_{p\g}=0
 \ee

Generally we can define a bilinear form
 \be
 <X_{ M },X_{ N }>=X_{ M }X_{ N }-(-1)^{g(X_{ M })g(X_{ N })}X_{ N }X_{ M }=C^{P}_{NM}X_{P}
 \ee
where $X$ can be either $T$ or $Q$ and $P=1,...,d+D$ (say the
first $d$ are $T$'s and the rest $D$ are $Q$'s). $g(X_{M})$ is the
Grassmann grading, $g(T)=0$ and $g(Q)=1$ and $C^{P}_{NM}$ are the
structure constants. The latter satisfy the graded antisymmetry
property
 \be
 C^{P}_{NM}=-(-1)^{g(X_{ M }) g(X_{ N })}C^{P}_{MN}
 \ee

We define the super-metric on the super-algebra as the supertrace
of the generators in the fundamental representation
 \be
 g_{MN}=\Str X_M X_N,\label{supermetric}
 \ee
We can further define raising and lowering rules when the metric
acts on the structure constants
 \be
C_{ M  N P}\equiv g_{ M  S }C^{ S }_{ N P}
 \ee
 \be
C_{ M  N P}=-(-1)^{g(X_{ N })g(X_{P})}C_{ M PN }=-(-1)^{g(X_{ M
})g(X_{ N })}C_{ N  M P}
 \ee
 \be
C_{ M  N P}=-(-1)^{g(X_{ M })g(X_{ N })+g(X_{ N
})g(X_{P})+g(X_{P})g(X_{ M })}C_{P N  M }
 \ee
For a semi-simple super Lie algebra ($|g_{ M  N }|\neq0$ and
$|h_{mn}|\neq0$) we can define a contravariant metric tensor
through the relation
 \be
g_{ M P}g^{PN }=\d^{ N }_{ M }
 \ee

The Killing form is defined as the supertrace of the generators in
the adjoint representation
 \be
K_{ M  N }\equiv (-1)^{g(X_{P})}C^{ S }_{P M }C^{P}_{ S  N
}=(-1)^{g(X_{ M })g(X_{ N })}K_{ N  M }
 \ee
(while on the (sub)Lie-algebra we define the metric
$K_{mn}=f^{p}_{mq}f^{q}_{np}$). Explicitly we have
 \be
K_{mn}=h_{mn}-F^{\b}_{m\a}F^{\a}_{n\b}=K_{nm}
 \ee
 \be
K_{\a\b}=F^{\g}_{m\a}A^{m}_{\b\g}-F^{\g}_{m\b}A^{m}_{\a\g}=-K_{\b\a}
 \ee
 \be
K_{m\a}=K_{\a m}=0
 \ee
The Killing form is proportional to the supermetric up to the
second Casimir $C_2(G)$ of the supergroup, which is also called
the dual Coxeter number
 \be
 K_{MN}=-\,C_2(G)\,g_{MN}.
 \ee

In the main text, we have computed the one-loop beta-functions in
the background field method. It turns out that the sums of one-loop
diagrams with fixed external lines are proportional to the Ricci
tensor $R_{MN}$ of the supergroup. The super Ricci tensor of a
supergroup is defined as
 \bea\label{superi}
 R_{MN}(G)&=&-{1\over 4} f^P_{MQ}f^{Q}_{NP}(-)^{g(X_Q)},
 \eea
and we immediately see that $R_{MN}=-K_{MN}$, in particular, we
can write it as
 \bea
 R_{MN}(G)&=&{C_2(G)\over4}g_{MN},
 \eea

\subsection{Summary of our models}

We would like to have a clear spacetime interpretation of the dual
Coxeter number of a supergroup. Let us consider a supergroup $G$
with a $\ZZ_4$ automorphism, whose zero locus we denote by $H$. The
various RR backgrounds we discussed in the main text are realized
as $G/H$ supercosets of this kind. The bosonic submanifold is in
general $AdS_p\times S^q$, where the gauge group
$H=SO(1,p-1)\times SO(q)\times SO(r)$, and the $SO(r)$ factor
corresponds to the non-geometric isometries. We have the following
cases
 $$
 \begin{array}{cccccccccc}
 &    G     & {\rm Algebra} &&p  &  q  &  r &&\#_{susy} &C_2(G)\\
AdS_2 &\, Osp(1|2) & \,B(0|1)& & 2  &  0  &  0&&2&-3\\
AdS_2 &\, Osp(2|2) &  \,C(2)  && 2& 0  &  2&&4&-2\\
AdS_4 &\,Osp(2|4)  & \,C(3) && 4  &  0  &  2&&8&-4\\
AdS_6 & \,F_4      &\,  F(4;3) &&6  &  0  & 3 && 16&2\\
AdS_5\times S^1&\,SU(2,2|2)& \,A(2|4)& &  5  &  0  &  3&&16&4\\
AdS_2\times S^2& \,PSU(1,1|2) &\,A(2|2)&&  2  & 2  &  0& &8&0\\
AdS_3\times S^3&   \,PSU(1,1|2)^2&\,A(2|2)\oplus A(2|2) &&3  & 3  &  0&& 16&0\\
AdS_5\times S^5&\,PSU(2,2|4)&\,A(4|4)& &  5  & 5  &  0&& 32&0
\label{tablespaces}
\end{array}
 $$
The superspace notations will be as follows: the letters
$\{M,N,\ldots\}$ refer to elements of the supergroup $G$, while
$\{I,J,\ldots\}$ take values in the gauge group $H$ and finally
$\{A,B,\ldots\}$ refer to elements of the supercoset $G/H$. The
lower case letters denote the bosonic and fermionic components of
the superspace indices, while $\#_{susy}$ is the number of real
spacetime supercharges in the background. Then, we can rewrite the
super Ricci tensor of the supergroup (\ref{superi}) making
explicit the $\ZZ_4$ grading\footnote{The Ricci tensor of the
supercoset $G/H$ is given $R_{AB}(G/H)=-{1\over 4}
f^C_{AD}f^{D}_{BC}(-)^{C}-f^I_{AD}f^D_{BI}(-)^{I}$, see
\cite{Berkovits:1999zq}.}
 \bea
 R_{AB}(G)&=&-{1\over 4} f^C_{AD}f^{D}_{BC}(-)^C-\half f^I_{AD}f^D_{BI}(-)^I,
 \eea
In particular, its grading two part is
 \bea
 R_{ab}(G)={1\over4}\left(F^\a_{a\hat\b}F^\bh_{b\a}+F^\ah_{a\a}F^\a_{b\ah}\right)-\half
 f^{i}_{ac}f^{c}_{bi},
 \eea

\subsection{$Osp(2|2)$}

The \textrm{Osp(2$\vert$2)} supergroup corresponds to the
superalgebra $C(2)$. It has a bosonic subgroup $Sp(2)\times SO(2)$
and four real fermionic generators transforming in the ${\bf
4}\oplus{\bf 4}$ of $Sp(2)$. It consists of the super matrices
$\mathbf{M}$ satisfying $\mathbf{M^{st}HM=H}$, where
\begin{displaymath}
\mathbf{H} = \left( \begin{array}{cc|cc}
0 & 1 & 0 & 0 \\
-1 & 0 & 0 & 0 \\
\hline
0 & 0 & 1 & 0 \\
0 & 0 & 0 & 1
\end{array} \right)
\end{displaymath}
The superalgebra is obtained by the commutation relations
$\mathbf{m^{st}H+Hm}=0$, where we parameterize \bea
\begin{array}{ccc} \mathbf{m} &=& \left( \begin{array}{cc|cc}
sl(2) & {} & a & b \\
{} & {} & c & d \\
\hline
e & f & {} & {} \\
g & h & {} & so(2)
\end{array} \right)
\end{array}
&\qquad
\begin{array}{ccc}
\mathbf{m^{st}} &=& \left( \begin{array}{cc|cc}
sl(2)^{t} & {} & e & g \\
{} & {} & f & h \\
\hline
-a & -c & {} & {} \\
-c & -d & {} & so(2)^{t}
\end{array} \right)
\end{array}
 \eea
so that from the condition $\mathbf{m^{st}H+Hm} =0$ we find
\begin{displaymath}
\mathbf{m} = \left( \begin{array}{cc|cc}
sl(2) & {} & a & b \\
{} & {} & c & d \\
\hline
-c & a & {} & {} \\
-d & b & {} & so(2)
\end{array} \right).
\end{displaymath}
The Cartan basis for the $Osp(2|2)$ superalgebra is given by the
following supermatrices. The bosonic generators are
\begin{displaymath}
\bf H= \left(
\begin{array}{cc|cc}
1 & 0 & {} & {} \\
0 & -1 & {} & {} \\
\hline
{} & {} & {} & {} \\
{} & {} & {} & {}
\end{array} \right),
\qquad \bf E^{+}= \left(
\begin{array}{cc|cc}
0 & 1 & {} & {} \\
0 & 0 & {} & {} \\
\hline
{} & {} & {} & {} \\
{} & {} & {} & {}
\end{array} \right),
\qquad \bf E^{-}= \left(
\begin{array}{cc|cc}
0 & 0 & {} & {} \\
1 & 0 & {} & {} \\
\hline
{} & {} & {} & {} \\
{} & {} & {} & {}
\end{array} \right),
\qquad \bf \widetilde{H}= \left(
\begin{array}{cc|cc}
{} & {} & {} & {} \\
{} & {} & {} & {} \\
\hline
{} & {} & 0 & 1 \\
{} & {} & -1 & 0
\end{array} \right),
\end{displaymath}
where $(\bf H,\bf E^\pm)$ are the generators of $sl(2)$ while $\bf
\tilde H$ is the generator of $SO(2)$. The fermionic generators
$(\bf Q_\a,\bf Q_\ah)$ are
\begin{displaymath}
\bf Q_1= \frac{1}{\sqrt{2}}\left(
\begin{array}{cc|cc}
{} & {} & 0 & 1 \\
{} & {} & 0 & 0 \\
\hline
0 & 0 & {} & {} \\
0 & 1 & {} & {}
\end{array} \right),
\qquad \bf Q_{\hat 1}= \frac{1}{\sqrt{2}}\left(
\begin{array}{cc|cc}
{} & {} & 0 & 0 \\
{} & {} & 0 & -1 \\
\hline
0 & 0 & {} & {} \\
1 & 0 & {} & {}
\end{array} \right),
\end{displaymath}
\begin{displaymath}
\bf Q_2= \frac{1}{\sqrt{2}}\left(
\begin{array}{cc|cc}
{} & {} & 1 & 0 \\
{} & {} & 0 & 0 \\
\hline
0 & 1 & {} & {} \\
0 & 0 & {} & {}
\end{array} \right)
\qquad \bf Q_{\hat 2}= \frac{1}{\sqrt{2}}\left(
\begin{array}{cc|cc}
{} & {} & 0 & 0 \\
{} & {} & -1 & 0 \\
\hline
1 & 0 & {} & {} \\
0 & 0 & {} & {}
\end{array} \right)
\end{displaymath}
\\
Finally, the $Osp(2|2)$ superalgebra is given by
 \bea
[\HH,\EE^\pm]=\pm 2\EE^\pm,& [\EE^+,\EE^-]=\HH
,& [\HH,\tilde \HH]=0,\nonumber\\
\,[ \tilde \HH, \EE^\pm ]=0, & [\tilde
\HH,\QQ_\a]=\epsilon_{\a\b}\QQ_\b,& [\tilde
\HH,\QQ_\ah]=\epsilon_{\ah\hat\b}\QQ_{\hat \b},\nn\\
\, [\HH,\QQ_\a]=  \QQ_\a,& [
\HH,\QQ_\ah]=-\QQ_{\hat \a},\label{osp2}\\
\{\QQ_\a,\QQ_\b\}=\half\delta_{\a\b}\EE^+, &
\quad\{\QQ_\ah,\QQ_{\hat
\b}\}=\half\delta_{\ah\hat\b}\EE^-,&\quad\{\QQ_\a,\QQ_\ah\}=\half\delta_{\a\ah}
\HH+\half\epsilon_{\a\ah}\tilde \HH,\nn
\\
\,[\EE^+,\QQ_\a]=[\EE^-,\QQ_\ah]=0,&[\EE^+,\QQ_\ah]=-\delta_{\ah\a}\QQ_\a,&
[\EE^-,\QQ_\a]=-\delta_{\a\ah}\QQ_\ah,\nn
 \eea
We classify the generators according to their $\ZZ_4$ charge
 \be
 \begin{array}{c|cc|cc|cc}
 {\cal H}_0 &{}\,& {\cal H}_1 & {}\,&{\cal H}_2 &{}\,& {\cal H}_3 \\
 &&&&&&\\
 \HH,\tilde \HH &{}& \QQ_\a &{}& \EE^\pm &{}& \QQ_\ah.
  \end{array}
 \ee
In the main text, we realize our $AdS_2$ background by quotienting
with respect to the grading zero subgroup, namely $SO(1,1)\times
SO(2)$. The structure constants are
 \be
 f_{ml}^H=\delta_m^+\delta_l^--\delta_m^-\delta_l^+,\quad
 f_{ml}^{\tilde H}=0,
 \ee
$$f_{Hm}^l=2(\delta_m^+\delta^l_+-\delta_m^-\delta^l_-),\qquad f_{\tilde Hm}^l=0$$
$$F_{\a m}^\ah=\delta_\a^\ah\delta_m^-,\qquad
  F_{\ah m}^\a=\delta_\ah^\a\delta_m^+ $$
$$F_{H\a}^\b=\delta_\a^\b,\qquad
  F_{H\ah}^{\hat \b}=-\delta_\ah^{\hat\b}$$
$$F_{\tilde H\a}^\b=\epsilon_{\a\gamma}\delta^{\gamma\b},\qquad
  F_{\tilde H\ah}^{\hat\b}=\epsilon_{\ah\hat\gamma}\delta^{\hat\gamma\hat\b}$$
$$A_{\a\b}^m=\delta_{\a\b}\delta^{m}_+,\qquad
  A_{\ah\hat\b}^m=-\delta_{\ah\hat\b}\delta^{m}_-,$$
$$A^H_{\a\hat\b}=A^H_{\hat\b\a}=\half\delta_{\a\hat\b},\qquad
  A^{\tilde H}_{\a\hat\b}=A^{\tilde H}_{\hat\b\a}=\half\epsilon_{\a\hat\b},$$

The metric on the supergroup is
 \bea
 g_{mn}&=&\delta_{m}^{+}\delta_{n}^{-}+\delta_{m}^{-}\delta_{n}^{+},\\
 g_{\a\ah}&=&-\eta_{\ah\a}=\delta_{\a\ah},\qquad
 g_{ij}=2\delta_{ij},\nn
 \eea
where $m,n=\pm$, $i,j=H,\tilde H$.

The $OSp(1|2)$ supergroup corresponds to the superalgebra
$B(0|1)$. Its bosonic subgroup is $Sp(2)$ and it has two real
fermionic generators transforming in the ${\bf 2}$ of $Sp(2)$. It
can be easily obtained by the one of the $Osp(2|2)$ supergroup by
simply dropping the generators $\bf \widetilde H$ and
$\QQ_2,\QQ_{\hat 2}$.

\subsection{$Osp(2|4)$}

The supergroup $Osp(2|4)$ corresponds to the superalgebra $C(3)$.
Its bosonic subgroup is $Sp(4)\times SO(2)$ and it has eight real
fermionic generators transforming in the ${\bf 4}\oplus{\bf 4}$ of
$Sp(4)$. We classify the generators according to their $\ZZ_4$
charge
 \be
 \begin{array}{c|cc|cc|cc}
 {\cal H}_0 &{}\,& {\cal H}_1 & {}\,&{\cal H}_2 &{}\,& {\cal H}_3 \\
 &&&&&&\\
 {\bf J_{[ab]}},\tilde \HH &{}& \QQ_\a &{}& {\bf P}_a &{}&
 \QQ_\ah,
  \end{array}
 \ee
where $a=0,\ldots,3$ and $\a,\ah$ are four-dimensional Majorana
spinor indices. In the main text, we realize our $AdS_4$
background by quotienting with respect to the grading zero
subgroup, namely $SO(1,3)\times SO(2)$. The structure constants
are
  \be f_{a b}^{[cd]}=
\half \d_{a}^{[c} \d_b^{d]},\quad
f_{a[bc]}^{d}=-f_{[bc]a}^{d}=\eta_{a[b}\d^{d}_{c]}\ee
$$f_{[ab][cd]}^{[ef]}=\frac{1}{2}\d^{[e}_{[a}\eta_{b][c}\d^{f]}_{d]}
=\frac{1}{2}\left(
 \eta_{bc}\d^{[e}_{a}\d^{f]}_{d}+\eta_{ad}\d^{[e}_{b}\d^{f]}_{c}
-\eta_{ac}\d^{[e}_{b}\d^{f]}_{d}-\eta_{bd}\d^{[e}_{a}\d^{f]}_{c}
\right)$$
$$F^{\bh}_{a\a}=-F^{\bh}_{\a a}=\frac{1}{2}(\g_{a})^{\b}{}_{\a}\d^{\bh}{}_{\b}, \quad
F^{\b}_{a\ah}=-F^{\b}_{\ah
a}=\frac{1}{2}(\g_{a})^{\bh}{}_{\ah}\d^{\b}{}_{\bh}$$
$$F^{\b}_{[ab]\a}=-F^{\b}_{\a[ab]}=\frac{1}{2}(\g_{ab})^{\b}{}_{\a}, \quad
F^{\bh}_{[ab]\ah}=-F^{\bh}_{\ah[ab]}=\frac{1}{2}(\g_{ab})^{\bh}{}_{\ah}$$
$$F_{\tilde H\a}^{\b}=\frac{1}{2}(\g^{5})_{\a}{}^{\b}, \quad
F_{\tilde H\ah}^{\bh}=-\frac{1}{2}(\g^{5})_{\ah}{}^{\bh}$$
$$A^{a}_{\a\b}=(C\g^{a})_{\a\b}, \quad
A^{a}_{\ah\bh}=(C\g^{a})_{\ah\bh}$$
$$A^{\tilde H}_{\a\bh}=-2(\g^{5})_{\a}{}^{\g}(\tilde{C})_{\g\bh}, \quad
A^{\tilde H}_{\ah\b}=2(\g^{5})_{\ah}{}^{\gh}(\tilde{C})_{\gh\b}$$
$$A^{[ab]}_{\a\bh}=-\frac{1}{2}(\tilde{C})_{\a\gh}(\g^{ab})^{\gh}{}_{\bh}, \quad
A^{[ab]}_{\ah\b}=-\frac{1}{2}(\tilde{C})_{\ah\g}(\g^{ab})^{\g}{}_{\b}$$
where $C$ is the charge conjugation matrix of $SO(1,3)$. The
supermetric is given by
 \bea
 g_{ab}&=&\eta_{ab},\qquad g_{\a\bh}=2C_{\a\bh}\\
 g_{[ab][cd]}&=&\eta_{a[c}\eta_{d]b},\qquad g_{\tilde H\tilde H}=2.\nonumber
 \eea

\subsection{$SU(2,2|2)$}

The supergroup $SU(2,2|2)$ corresponds to the superalgebra
$A(3|1)$. Its bosonic subgroup is $SU(2,2)\times SO(3)\times U(1)$
and it has sixteen real fermionic generators transforming in the
$({\bf \bar4},{\bf 2})\oplus({\bf 4},{\bf \bar2})$. We classify
the generators according to their $\ZZ_4$ charge
 \be
 \begin{array}{c|cc|cc|cc}
 {\cal H}_0 &{}\,& {\cal H}_1 & {}\,&{\cal H}_2 &{}\,& {\cal H}_3 \\
 &&&&&&\\
 {\bf J_{[ab]}},{\bf \tilde H_{a'}} &{}& \QQ_{\a\a'} &{}& {\bf P}_a,{\bf R} &{}&
 \QQ_{\ah\ah'},
  \end{array}
 \ee
where $a=0,\ldots,4$ are the coordinates on the $AdS_5$, ${ \bf
R}$ is the translation generator on $S^1$ and $a'=1,2,3$ is the
$SO(3)$ vector index, while $(\a,\ah)$ are Majorana spinor indices of
$SO(1,4)$ and $(\a',\ah')$ are spinors of $SO(3)$. In the main
text, we realize our $AdS_5\times S^1$ background by quotienting
with respect to the grading zero subgroup.

Its structure constants are
  \be
 f_{a b}^{[cd]}= \half \d_{a}^{[c}\d_b^{d]},\quad f_{a' b'}^{c'}=
 \e_{a'b'}{}^{c'},\quad f_{ {a}[ {bc}]}^{ {d}}=-f_{[ {bc}] {a}}^{ {d}}=\eta_{ {a}[
{b}}\d^{ {d}}_{ {c}]}
 \ee
$$f_{[ {ab}][ {cd}]}^{[ {ef}]}=\frac{1}{2}\d^{[ {e}}_{[ {a}}\eta_{ {b}][ {c}}\d^{ {f}]}_{ {d}]}
=\frac{1}{2}\left(
 \eta_{ {bc}}\d^{[ {e}}_{ {a}}\d^{ {f}]}_{ {d}}+\eta_{ {ad}}\d^{[ {e}}_{ {b}}\d^{ {f}]}_{ {c}}
-\eta_{ {ac}}\d^{[ {e}}_{ {b}}\d^{ {f}]}_{ {d}}-\eta_{ {bd}}\d^{[
{e}}_{ {a}}\d^{ {f}]}_{ {c}} \right)$$
$$F^{\bh\bh'}_{\a\a'a}=-\frac{i}{2}(\g_{a})^{\bh}{}_{\a}\d^{\bh'}{}_{\a'},\qquad
  F^{\b\b'}_{\ah\ah'a}= \frac{i}{2}(\g_{a})^{\b}{}_{\ah}\d^{\b'}{}_{\ah'}$$
$$F^{\bh\bh'}_{\a\a'R}= \frac{1}{2}\d^{\bh'}{}_{\a'}\d^{\bh}{}_{\a},\qquad
  F^{\b\b'}_{\ah\ah'R}=-\frac{1}{2}\d^{\b'}{}_{\ah'}\d^{\b}{}_{\ah}$$
$$F^{\b\b'}_{\a\a'[ ab]}=-\frac{1}{2}(\g_{ {ab}})^{\b}{}_{\a}\d^{\b'}{}_{\a'},\qquad
  F^{\bh\bh'}_{\ah\ah'[ ab]}= -\frac{1}{2}(\g_{ {ab}})^{\bh}{}_{\ah}\d^{\bh'}{}_{\ah'}$$
$$F^{\b\b'}_{\a\a' a'}=-\frac{1}{2}(\tau_{a'})^{\b'}{}_{\a'}\d^{\b}{}_{\a},\qquad
  F^{\bh\bh'}_{\ah\ah' a'}= -\frac{1}{2}(\tau_{a'})^{\bh'}{}_{\ah'}\d^{\bh}{}_{\ah}$$
$$A^{a}_{\a\a'\b\b'}=-iC_{\a'\b'}(C\g^{a})_{\a\b},\qquad
  A^{a}_{\ah\ah'\bh\bh'}=-iC_{\ah'\bh'}(C\g^{a})_{\ah\bh}$$
$$A^{R}_{\a\a'\b\b'}=C_{\a\b}C'_{\a'\b'},\qquad
  A^{R}_{\ah\ah'\bh\bh'}=C_{\ah\bh}C'_{\ah'\bh'}$$
$$A^{[ab]}_{\a\a'\bh\bh'}=\half C_{\a'\bh'}(C\g^{ab})_{\a\bh},\qquad
  A^{[ab]}_{\ah\ah'\b\b'}=-\half C_{\ah'\b'}(C\g^{ab})_{\ah\b}$$
$$A^{a'}_{\a\a'\bh\bh'}=-2C_{\a\bh}(C'\tau^{a'})_{\a'\bh'},\qquad
  A^{a'}_{\ah\ah'\b\b'}= 2C_{\ah\b}(C'\tau^{a'})_{\ah'\b'},$$
where $C_{\a\b}$ and $C'_{\a'\b'}$ are the charge conjugation
matrices respectively of $SO(1,4)$ and $SO(3)$. The supermetric is
 \bea
 g_{ab}&=&-\eta_{ab},\qquad  g_{RR}=1,\\
 g_{\a\bh}&=&C_{\a\bh},\qquad g_{\a'\bh'}=C'_{\a'\bh'}.\nonumber
 \eea

\subsection{$AdS_p\times S^p$ superalgebras}
\label{adspsp}

The $AdS_p\times S^p$ backgrounds are realized by the following
supercosets
 \be
 \begin{array}{ccccc}
 AdS_2\times S^2 &{}\,& AdS_3\times S^3 &{}\,& AdS_5\times S^5\\
 {PSU(1,1|2)\over SO(1,1)\times SO(2)}&{}\,&{PSU(1,1|2)
^2\over SO(1,2)\times
 SO(3)}&{}\,&{PSU(2,2|4)\over SO(1,4)\times SO(5)}
 \end{array}
 \ee
We can treat the supergroups $PSU(1,1|2)$, $PSU(1,1|2)^2$ and
$PSU(2,2|4)$ schematically altogether, by collecting their
generators according to their $\ZZ_4$ grading as follows
 \be
 \begin{array}{c|cc|cc|cc}
 {\cal H}_0 &{}\,& {\cal H}_1 & {}\,&{\cal H}_2 &{}\,& {\cal H}_3 \\
 &&&&&&\\
 {\bf J_{[ab]}},{\bf J_{a'b'}} &{}& \QQ_{\a\a'} &{}& {\bf P}_a,{\bf P}_{a'} &{}&
 \QQ_{\ah\ah'},
  \end{array}
 \ee
where $a=0,\ldots,p-1$ are the coordinates on the $AdS_p$,
$a'=1,\ldots,p$ are the coordinates along $S^p$, while $(\a,\ah)$
are Weyl spinor indices of $SO(1,p-1)$ and $(\a',\ah')$ are
spinors of $SO(p)$. In the main text, we realize our $AdS_p\times
S^p$ backgrounds by quotienting with respect to the grading zero
subgroup. The structure constants are
 \be
 f_{a b}^{[cd]}= \half \d_{a}^{[c}\d_b^{d]}\quad f_{a' b'}^{[c'd']}= -\half \d_{a'}^{[c'}\d_b'^{d']}
 \quad f_{\underline{a}[\underline{bc}]}^{\underline{d}}=
-f_{[\underline{bc}]\underline{a}}^{\underline{d}}=
\eta_{\underline{a}[\underline{b}}\d^{\underline{d}}_{\underline{c}]}\ee
$$f_{[\underline{ab}][\underline{cd}]}^{[\underline{ef}]}=
\frac{1}{2}\d^{[\underline{e}}_{[\underline{a}}\eta_{\underline{b}]
[\underline{c}}\d^{\underline{f}]}_{\underline{d}]}
=\frac{1}{2}\left(
 \eta_{\underline{bc}}\d^{[\underline{e}}_{\underline{a}}\d^{\underline{f}]}_{\underline{d}}
 +\eta_{\underline{ad}}\d^{[\underline{e}}_{\underline{b}}\d^{\underline{f}]}_{\underline{c}}
-\eta_{\underline{ac}}\d^{[\underline{e}}_{\underline{b}}\d^{\underline{f}]}_{\underline{d}}
-\eta_{\underline{bd}}\d^{[\underline{e}}_{\underline{a}}\d^{\underline{f}]}_{\underline{c}}
\right)$$
$$F^{\bh\bh'}_{\a\a'a}=-\frac{i}{2}(\g_{a})^{\bh}{}_{\a}\d^{\bh'}{}_{\a'},\qquad
  F^{\b\b'}_{\ah\ah'a}= \frac{i}{2}(\g_{a})^{\b}{}_{\ah}\d^{\b'}{}_{\ah'}$$
$$F^{\bh\bh'}_{\a\a'a'}= \frac{1}{2}(\g_{a'})^{\bh'}{}_{\a'}\d^{\bh}{}_{\a},\qquad
  F^{\b\b'}_{\ah\ah'a'}=-\frac{1}{2}(\g_{a'})^{\b'}{}_{\ah'}\d^{\b}{}_{\ah}$$
$$F^{\b\b'}_{\a\a'[\underline{ab}]}=-\frac{1}{2}(\g_{\underline{ab}})^{\b}{}_{\a}\d^{\b'}{}_{\a'},\qquad
  F^{\bh\bh'}_{\ah\ah'[\underline{ab}]}= \frac{1}{2}(\g_{\underline{ab}})^{\bh}{}_{\ah}\d^{\bh'}{}_{\ah'}$$
$$A^{a}_{\a\a'\b\b'}=-iC_{\a'\b'}(C\g^{a})_{\a\b},\qquad
  A^{a}_{\ah\ah'\bh\bh'}=-iC_{\ah'\bh'}(C\g^{a})_{\ah\bh}$$
$$A^{a'}_{\a\a'\b\b'}=C_{\a\b}(C'\g^{a'})_{\a'\b'},\qquad
  A^{a'}_{\ah\ah'\bh\bh'}=C_{\ah\bh}(C'\g^{a'})_{\ah'\bh'}$$
$$A^{[ab]}_{\a\a'\bh\bh'}=\half C_{\a'\bh'}(C\g^{ab})_{\a\bh},\qquad
  A^{[ab]}_{\ah\ah'\b\b'}=-\half C_{\ah'\b'}(C\g^{ab})_{\ah\b}$$
$$A^{[a'b']}_{\a\a'\bh\bh'}=-\half C_{\a\bh}(C'\g^{a'b'})_{\a'\bh'},\qquad
  A^{[a'b']}_{\ah\ah'\b\b'}= \half C_{\ah\b}(C'\g^{a'b'})_{\ah'\b'}$$
where $C_{\a\b}$ and $C'_{\a'\b'}$ are respectively the charge
conjugation matrices of $SO(1,p-1)$ and $SO(p)$. The supermetric
in the fundamental is
 \bea
 g_{ab}&=&-\eta_{ab},\qquad g_{a'b'}=\eta_{a'b'},\\
 g_{\a\ah}&=& C_{\a\ah}C'_{\a'\ah'},
 \eea

\section{Non-critical supergravity in $d$ dimensions}
\label{sixsugra}

In this section we study $d$-dimensional supergravity with a
cosmological constant. We will show that there is an
$AdS_{d-1}\times S^1$ solution with RR $(d-1)$-form flux only when
we introduce space-filling sources, that can be interpreted as
uncharged $D_{d-1}$-branes. This reproduces the results found in
\cite{Klebanov:2004ya} for the $AdS_5\times S^1$ case and gives
the new solution $AdS_3\times S^1$. In the next section we will
then argue that, by including the first $\a'$ corrections to the
non-critical supergravity equations, an $AdS_5\times S^1$ solution
might be possible, even without the space-filling sources.

The $d$-dimensional non-critical supergravity action in the string
frame is
 \bea
 S&=&{1\over \kappa^2}\int d^dx\,\sqrt{-G}
 \left[-(\partial_\mu\chi)^2+e^{-2\phi}\left(R+4(\partial_\mu\phi)+\Lambda\right)
 -2N_fe^{-\phi}\right],\label{sugraaction}
 \eea
where $\chi$ is the RR scalar, dual to the $(d-1)$-form flux, the
cosmological constant is
 \bea
 \Lambda={10-d\over \alpha'},
 \eea
and the last term is the contribution of $N_f$ pairs of
space-filling uncharged sources. In \cite{Klebanov:2004ya} this
term was interpreted as arising from $N_f$ pairs of branes and
anti-branes. We make an ansatz for a solution of the kind
$AdS_{d-1}\times S^1$ with constant dilaton. The equations of
motion for the metric and the dilaton then reduce to
 \bea
 R_{\mu\nu}&=&{1\over D-2}
 G_{\mu\nu}\left(2N_fe^\phi-\Lambda\right)+e^{2\phi}\partial_\mu\chi\partial_\nu\chi
 ,\label{sugra6}\\
0&=&R+\Lambda-N_fe^\phi, \nn
 \eea
Let us parameterize the dual of the RR $(d-1)$-form flux as a
one-form flux $\partial_\mu\chi$, whose only nonzero leg is along
the $S^1$ as in the previous section. In particular, $\chi\sim
N_c\theta$, where $\theta$ is the coordinate on the circle, so
that $\partial_\theta\chi\sim N_c$. Note that component of the
Ricci curvature along the circle vanishes $R_{\t\t}=0$. Then we
find the following solution for the scalar curvature and the
string coupling
 \bea
 R=-{10-d\over d}(d-1),&\qquad& g_S\equiv e^\phi={10-d\over d}N_f,
 \eea
and the components of the Ricci curvature along the $AdS_{d-1}$
are $R_{ij}=-{10-d\over d}G_{ij}$. Recalling that
$R_{ij}=-{d-2\over R_{AdS}^2}G_{ij}$ and $G_{\t\t}=R_{S}^2$ we can
read off the radii
 \bea
 R_{AdS}^2={d(d-2)\over 10-d}\a',&\qquad& R_{S}^2={10-d \over d}{N_c^2\over N_f^2}.
 \eea
It is easy to see that, without space-filling sources, namely if we
set $N_f=0$, then there is no solution to the supergravity
equations (\ref{sugra6}). In the case $d=6$ we recover the
$AdS_5\times S^1$ solution of \cite{Klebanov:2004ya}. Moreover, we
find the new solution
 \be
 AdS_3\times S^1,\qquad R_{AdS}^2={4\over 3}\alpha',\qquad
 R_S^2={3\over2}{N^2\over N_f^2}\alpha'.
 \ee
It would be interesting to repeat the analysis of
\cite{Klebanov:2004ya} for this last case, to understand its
relation to the four-dimensional type II linear dilaton
background.

\subsection{Higher curvature corrections}

We have just seen above that there is no $AdS_5\times S^1$
solution to the six-dimensional non-critical supergravity
equations, unless we include space-filling sources. We will now
argue that, when we include the first $\a'$ corrections to the
supergravity equations, there might be a solution {\it without}
the space-filling sources. At the end of the section, we will
comment about the validity of our argument.

We use the methods of \cite{Kuperstein:2004yk} and \cite{cobi}. We
make the ansatz for a solution of non-critical supergravity of the
form $AdS_5\times S^1$ with constant dilaton $g_s=e^\phi$ and
constant RR five-form flux $F_5$. We are not adding any
space-filling brane. Let us denote by $R_{AdS},R_S$ the radii of
$AdS$ and $S^1$ respectively and parameterize the dual of the RR
five-form flux as a one form flux $\partial_\mu\chi$, whose only
nonzero leg is along the $S^1$ as in the previous section, in
particular $\chi\sim N_c\theta$. If we plug these ansatze in the
ordinary supergravity action (\ref{sugraaction}), we find the
leading order terms
 \bea
 S_0&=&V g_s^{-2}R_{AdS}^2R_S\left(-{20\over R_{AdS}^2}+\Lambda-{(g_sN)^2\over
 R_S^2}\right),
 \eea
where $\Lambda=(10-d)/\alpha'$. Note that all the components of
the Riemann tensor along the circle direction vanish, so that we
do not have a term proportional to $R_S^{-2}$, which would come
from the ordinary Einstein-Hilbert action.

Let us address now the first $\a'$ corrections to the supergravity
action, which are basically of three kinds. The curvature squared
terms of the kind $R_{\mu\nu\rho\sigma}^2,R_{\mu\nu}^2,R^2$ are
all proportional to $R_{AdS}^{-4}$. The fourth power of the RR
field strength \bea
 (g^{\mu\nu}\partial_\mu\chi\partial\nu\chi)^2\sim {(g_sN_c)^4\over
 R_S^4},
 \eea
 while the only mixed coupling between the RR field strength and
 the curvatures is
 \bea
 R (g^{\mu\nu}\partial_\mu\chi\partial_\nu\chi)\sim {(g_sN_c)^2\over
 R_{AdS}^2R_S^2}.
 \eea
 Collecting all the terms we find the first $\alpha'$ corrections
 to the action
\bea
 S_1=V g_s^{-2} R_{AdS}^5R_S\left({\gamma\over R_{AdS}^4}+A{(g_sN_c)^2\over
 R_{AdS}^2R_S^2}+B{(g_sN_c)^4\over R_S^4}\right),
 \eea
where $\gamma,A,B$ are real coefficients.

The goal now is to vary the action $S_0+S_1$ with respect to
$R_{AdS},R_S,g_s$ and look for a real positive solution for some
values of the coefficients $A,B,\gamma$. The three variations of
the action read
 \bea
 {1\over 2B}\left({s\over
 a}\right)^2\left(-40a+8a^2+{2\gamma}\right)&=&g^2,\nn\\
 -20 s^2a+4s^2a^2+gsa^2+\gamma s^2-Agsa-3Bg^2a^2&=&0,\label{redusugra}\\
 -60 as^2+20a^2s^2-5a^2sg+\gamma s^2+3Agas+5Ba^2g^2&=&0.\nn
 \eea
where
$$a\equiv R_{AdS}^2,\quad s\equiv R_S^2,\quad g\equiv (g_sN)^2.
$$
One can first check that if $A,B=0$ there is no solution to the
equations of motion. This is the result of \cite{cobi} that, if we
include only the curvature squared terms and not the RR couplings,
there is still no $AdS_5\times S^1$ solution.

It turns out that there is a solution with nonvanishing
coefficients $\gamma,A,B$, for real positive radii squared and
string coupling, namely
 \bea
 g&=&1,\nn\\
 s&=&{a\over 4(2a-5)},\label{nextsol}\\
  a&=&{5\over2}+\half\sqrt{\gamma-25\over 16B-1},\nn
 \eea
 and two choices of coefficients, related by analytic continuation. The first is
 \bea
 B>{1\over 16},\quad\gamma>25,\\
 A={5\over2}-\half\sqrt{(16B-1)(\gamma-25)}.
 \eea
 The second is
 \bea
 B<{1\over 16},\quad\gamma<25,\\
 A={5\over2}+\half\sqrt{(16B-1)(\gamma-25)}.
 \eea
Note that we always have $a>5/2$ so that $s>0$ in (\ref{nextsol}).

We conclude that, even if $AdS_5\times S^1$ is not a solution to the
one-loop beta-function equations for Weyl invariance, when we consider
all the $\alpha'$ corrections to the next order, there might be a
solution. This fact points towards the possibility of having a two
loop conformal invariant non-critical superstring on $AdS_5\times
S^1$. However, we have not actually proven that this choice of the
coefficients solve the full supergravity equation.  To show this, one
would have to check that this solution satisfies also the usual
gravity constraints (namely the vanishing of the stress tensor), which
must be taken into account if we honestly consider the supergravity
equations of motion and not just their reduction
(\ref{redusugra}). Additionally, the coefficients $A$, $B$, and
$\gamma$ computed exactly using string theory may not fall withing the
specified range. Moreover, the non-critical supergravity does not
provide a consistent approximation to the type II non-critical
superstring, as we pointed out in section \ref{validity}. This
conjectured non-critical superstring would be dual to a
four-dimensional ${\cal N}=2$ superconformal field theory. Indeed, the
supersymmetries and the global symmetries match on the two sides.

\bibliography{PhD}
\bibliographystyle{JHEP}

\end{document}